\documentclass[twocolumn,twocolappendix]{aastex701}

\usepackage{bm}

\newcommand{\rhog}{\rho_\mathrm{g}}
\newcommand{\vvg}{\bm{v}_\mathrm{g}}
\newcommand{\dvv}{\Delta \bm{v}}
\newcommand{\Pg}{P_\mathrm{g}}
\newcommand{\Eg}{E_\mathrm{g}}
\newcommand{\taud}{\tau_\mathrm{d}}
\newcommand{\rhod}{\rho_\mathrm{d}}
\newcommand{\tstopd}{t_\mathrm{stop,d}}
\newcommand{\vvd}{\bm{v}_\mathrm{d}}

\begin{document}

\title{
A Split-Flux Method for Dust--Gas Two-Fluid Simulations from Strong to Weak Coupling
}

\author[orcid=0000-0002-2707-7548,sname='Iwasaki']{Kazunari Iwasaki}
\affiliation{Center for Computational Astrophysics, National Astronomical Observatory of Japan, Mitaka, Tokyo 181-8588, Japan}
\affiliation{Astronomical Science Program, The Graduate University for Advanced Studies (SOKENDAI), 2-21-1 Osawa, Mitaka, Tokyo 181-8588, Japan}
\email[show]{kazunari.iwasaki@nao.ac.jp}  

%

%
%
%
%

\begin{abstract}

We develop a new method for simulating pressureless fluids of neutral dust coupled to magnetohydrodynamic gas through dust--gas drag.
It has been pointed out that, in the strong-coupling regime, dust--gas multifluid methods suffer from artificial variations in the dust--to--gas mass ratio when the dust and gas are evolved with independent numerical fluxes.
This problem arises from inconsistencies between the dust and gas mass fluxes. In the strong-coupling limit, the dust--to--gas mass ratio is advected with the local gas velocity.
Recently, a new method based on the Harten--Lax--van Leer (HLL) solver was proposed.
This method constructs the dust mass flux so that it is consistent with the gas mass flux, and 
successfully reduces artificial variations in the dust--to--gas mass ratio.
Here, we develop a \texttt{Split-Flux} method that can be coupled to an HLLD gas solver.
Through various numerical experiments,
we show that the \texttt{Split-Flux} method suppresses artificial variations in the dust--to--gas mass ratio, 
while preserving the low numerical diffusivity of the HLLD solver.

\end{abstract}



\section{Introduction} \label{sec:introduction}

Dust grains affect the thermal, chemical, and dynamical evolution of interstellar and 
circumstellar gas. 
They contribute to heating and cooling processes for the gas, 
promote molecule formation on their surfaces, and exchange momentum with the gas through drag forces. 
The strength of dynamical coupling between dust and gas 
depends on grain size, gas density, and the dust--gas relative velocity. 
Dust and gas move at the same velocity in the strong-coupling limit, whereas 
they evolve independently in the weak-coupling limit.

Dust--gas dynamics is commonly modeled 
either by Lagrangian particle methods \citep{YoudinJohansen2007ApJ...662..613Y,BaiStone2010ApJS..190..297B,YangJohansen2016ApJS..224...39Y} or by 
Eulerian pressureless-fluid methods \citep{PaardekooperMellema2006A&A...453.1129P,Benitez-Llambay2019ApJS..241...25B,HuangBai2022ApJS..262...11H,Krapp2024ApJS..271....7K,Verrier2025A&A...701A.174V}. 
These methods have complementary advantages and limitations, depending on the degree of dust--gas
coupling.
In weakly coupled regimes, particle methods can naturally 
treat orbit crossing and multi-streaming trajectories, 
whereas pressureless-fluid methods have intrinsic limitations because they assume 
a single dust velocity at each position.

In strong-coupling regimes, however, 
both particle and pressureless-fluid methods face difficulties.
In this regime, the dust and gas should be transported nearly as a single fluid. 
For particle methods, the particles are advected approximately with the local gas velocity.
However, particles advected using an interpolated gas velocity do not necessarily follow the 
same discrete continuity equation as the finite-volume gas density.  
As a result,  the dust particle distribution can become inconsistent with the gas density \citep{Commercon_etal2023A&A...671A.128C}.  
\citet{Verrier2025A&A...701A.174V} pointed out that
the pressureless-fluid method also suffers from large errors in the dust--to--gas mass ratio when 
the dust and gas are evolved using independent numerical fluxes \citep{PaardekooperMellema2006A&A...453.1129P,HuangBai2022ApJS..262...11H}.

The difficulties facing pressureless-fluid methods in the strong-coupling limit
have been discussed in terms of spatial-resolution requirements for two-fluid 
methods.
Using a two-phase SPH formulation, 
\citet{LaibePrice2012MNRAS.420.2345L} showed that 
the characteristic stopping length, $c_\mathrm{s}\tstopd$, 
must be spatially resolved to avoid 
artificial overdamping, where $c_\mathrm{s}$ is the sound speed 
and $\tstopd$ is the dust stopping time.
More recently, \citet{Verrier2025A&A...701A.174V} suggested
a possible resolution criterion, $\Delta x < L_\mathrm{d}$,
where the recoupling length $L_\mathrm{d}\sim c_\mathrm{s}\tstopd$
is the distance over which a dust--gas drift generated
behind a shock is damped by drag.
They further suggested that the dust and gas fluxes computed 
independently can lead to inconsistency in dust and gas density transport 
when $\Delta x \gtrsim L_\mathrm{d}$.

In many astronomical applications, however, these resolution requirements 
are difficult to meet.
To recover consistent dust and gas transport without resolving $L_\mathrm{d}$, 
\citet{Verrier2025A&A...701A.174V} proposed \texttt{HLLgd}, in which 
the numerical fluxes for the dust are made consistent with those for the gas 
in unresolved strong-coupling regimes, while the method 
remains applicable across regimes ranging from strong to weak dust--gas coupling.
However, \texttt{HLLgd} also requires the gas variables to be updated using 
the Harten--Lax--van Leer (HLL) numerical fluxes \citep{HartenLaxvanLeer1983}.
This can be a limitation for MHD applications, where less diffusive solvers such as HLLD
\citep{MiyoshiKusano2005JCoPh.208..315M} are often preferred.

In this paper, we develop a Split-Flux formulation for pressureless dust fluids that can be 
combined with an HLLD gas solver and remains applicable from strong to weak coupling
within the pressureless-fluid approximation.
The method is designed to recover dust transport consistent with the gas mass flux when the dust--gas drift is unresolved, while reducing to a pressureless-dust flux when the drift is resolved.
The rest of this paper is organized as follows. 
In Section \ref{sec:method}, we describe the basic equations.
In Section \ref{sec:implementation}, we first describe existing implementations,
including \texttt{HLLgd}, and then introduce the new Split-Flux method.
The results of numerical experiments are described in Section \ref{sec:experiments}.
In Section \ref{sec:discussion}, we discuss the implications for streaming instability
and the extension to multiple dust species.
Section \ref{sec:summary} summarizes our findings.

\section{Equations and Numerical Framework}\label{sec:method}

We treat the dust component as a pressureless fluid coupled 
to the gas through drag.
The basic MHD equations of a dust--gas two-fluid system
are given by 
\begin{equation}
    \frac{\partial \rhog}{\partial t} + \bm{\nabla}\cdot (\rhog \vvg)=0,
\end{equation}
\begin{equation}
    \frac{\partial \rhog \vvg}{\partial t} + 
    \bm{\nabla}\cdot \left( P_\mathrm{tot} \bm{\mathrm{I}} + \rhog \vvg\otimes \vvg 
    - \bm{B}\otimes \bm{B}\right)
    =
    - K (\vvg - \vvd),
    \label{eomg}
\end{equation}
\begin{eqnarray}
 &&   \frac{\partial \Eg}{\partial t} +  
    \bm{\nabla}\cdot \left\{ (\Eg + P_\mathrm{tot})\vvg - (\vvg\cdot\bm{B})\bm{B} \right\} \nonumber \\
     && \hspace{1cm}= - K (\vvg - \vvd)\cdot\vvg +  K (\vvd - \vvg)^2,
     \label{eoe}
\end{eqnarray}
\begin{equation}
    \frac{\partial \bm{B}}{\partial t} 
    - \bm{\nabla}\times (\vvg \times \bm{B})=0,
\end{equation}
\begin{equation}
    \frac{\partial \rhod}{\partial t} + \bm{\nabla}\cdot (\rhod \vvd)=0,
\end{equation}
and 
\begin{equation}
    \frac{\partial \rhod \vvd}{\partial t} + 
    \bm{\nabla}\cdot \left( \rhod \vvd\otimes \vvd \right)
    =- K (\vvd - \vvg),
    \label{eomd}
\end{equation}
where 
$\rhog$ and $\vvg$ are the gas density and velocity, $\bm{B}$ is the magnetic field, 
$\Pg$ and $P_\mathrm{tot} = \Pg + \bm{B}^2/2$ are  the gas and total pressures, 
$\Eg = \rhog \vvg^2/2 + \Pg/(\gamma-1)  + \bm{B}^2/2$ is the gas total energy density, 
and $\gamma$ is the adiabatic index.
$\rhod$ and $\vvd$ are the dust density and dust velocity, respectively, and 
$\bm{\mathrm{I}}$ is the identity tensor.
The dust and gas components interact through collisions, and $K$ denotes
the drag coefficient.
The dust stopping time $\tstopd$ can be expressed in terms of $K$, 
\begin{equation}
    \tstopd = \frac{\rhod}{K}.
\end{equation}
The dust--to--gas mass ratio is defined as $f_\mathrm{dg}=\rhod/\rhog$.
For simplicity, the dust charge is neglected in this paper. The Lorentz force 
does not act on dust grains in Equation (\ref{eomd}).

The first term on the right-hand side of Equation~(\ref{eoe}) is the work done by the drag force on the gas, whereas the last term represents 
frictional heating due to dust--gas drag. 
In general, the dissipated kinetic energy can be partitioned between 
the internal energies of the dust and gas.  
In this work, because the dust is treated as a pressureless fluid 
and its internal energy is not evolved, we assume that all the 
dissipated kinetic energy is converted into gas internal energy.
The total energy of the dust and gas, $E_\mathrm{g}+\rhod \vvd^2/2$, is conserved.

We implement our dust-fluid module in Athena++ \citep{Stone2020ApJS..249....4S}.
We use either HLLE \citep{HartenLaxvanLeer1983,Einfeldt1988} or
HLLD \citep{MiyoshiKusano2005JCoPh.208..315M} as the MHD Riemann solver, 
depending on the dust solver.
We adopt the VL2 time integrator, which is a predictor-corrector midpoint method, with MUSCL reconstruction \citep{vanLeer1979JCoPh..32..101V}.  
In the default implementation of {Athena++}, the spatial accuracy is degraded to 
first order during the first substep, which advances the conserved variables from $t^n$ to 
$t^n+\Delta t/2$. 
Our numerical experiments show that this treatment can lead to artificial structures 
in the dust component. Thus, we use the MUSCL reconstruction 
in all substeps of the VL2 time integrator.

The drag terms in Equations (\ref{eomg}) and (\ref{eomd}) can be stiff and require
an extremely small time step if an explicit time integrator is used.
We adopt the VL2 implicit time integrator proposed by \citet{HuangBai2022ApJS..262...11H}.

\section{Numerical Fluxes for Pressureless Dust Fluids }\label{sec:implementation}

In this section, we review existing
dust fluxes and introduce a new Split-Flux formulation that recovers the
gas-carried dust transport in the strong-coupling limit while retaining
pressureless-dust transport when the dust--gas drift is resolved.

The numerical fluxes at each cell interface 
are computed from the left and right interface states. 
In this section, we focus on the numerical fluxes along the 
$x$-axis, $\bm{F}_\mathrm{d} = (\rhod v_{x\mathrm{d}},
\rhod v_{x\mathrm{d}}^2, \rhod v_{x\mathrm{d}}v_{y\mathrm{d}}, \rhod v_{x\mathrm{d}}v_{z\mathrm{d}})$ 
because extensions to the numerical fluxes along the $y$- and $z$-axes are trivial.
The physical quantities of the left and right states are denoted by the subscripts L and 
R, respectively.

Table \ref{tab:method} summarizes the methods considered in this work.

\subsection{Previously Proposed Dust Fluxes}\label{sec:dustflux_literature}
\subsubsection{Dust Solvers Independent of the Gas Solver: A Pressureless Riemann Solver}\label{sec:PM06}

The Riemann problem for a pressureless fluid was
investigated by \citet{LeVeque2004}.
The dust velocities, 
$v_{x\mathrm{d,L}}$ and
$v_{x\mathrm{d,R}}$, 
are two characteristic velocities of the 
dust Riemann problem when $v_{x\mathrm{d,L}}<v_{x\mathrm{d,R}}$.
A vacuum forms between the two characteristics.
When $v_{x\mathrm{d,L}}>v_{x\mathrm{d,R}}$,
the characteristics overlap, 
and the entropy solution contains
the so-called delta shock, which 
separates the left and right states and moves at a speed of 
$(\sqrt{\rho_\mathrm{d,L}}v_{x\mathrm{d,L}}
+ \sqrt{\rho_\mathrm{d,R}}v_{x\mathrm{d,R}})/(\sqrt{\rho_\mathrm{d,L}} + \sqrt{\rho_\mathrm{d,R}})$.
\citet{PaardekooperMellema2006A&A...453.1129P} 
adopt the solution of the pressureless Riemann problem 
to calculate the dust numerical fluxes.

\subsubsection{ Dust Solvers Independent of the Gas Solver: Collisionless Crossing, \texttt{HB22}}\label{sec:HB22}

\citet{HuangBai2022ApJS..262...11H} consider the collisionless behavior of dust particles 
when the velocities of the left and right states are directed 
toward the initial discontinuity.
Instead of considering the delta shock 
for $v_{x\mathrm{d,L}}>0$ and 
$v_{x\mathrm{d,R}}<0$, 
\citet{HuangBai2022ApJS..262...11H} use the sum of the numerical fluxes 
of the left and right states to mimic the crossing of collisionless dust streams.
They propose the following dust numerical flux:
\begin{equation}
    \bm{F}_\mathrm{d} = \left\{
    \begin{array}{ll}
         \bm{F}_\mathrm{d,L} & v_{x\mathrm{d,L}} \ge  0
                       \;\;\;\mathrm{and}\;\;\; v_{x\mathrm{d,R}}\ge 0\\
         \bm{F}_\mathrm{d,R} & v_{x\mathrm{d,L}} \le 0\;\;\;\mathrm{and}\;\;\; v_{x\mathrm{d,R}}\le 0\\
         \bm{0} & v_{x\mathrm{d,L}} < 0\;\;\;\mathrm{and}\;\;\; v_{x\mathrm{d,R}}>0\\
         \bm{F}_\mathrm{d,L}+\bm{F}_\mathrm{d,R} & v_{x\mathrm{d,L}} > 0\;\;\;\mathrm{and}\;\;\; v_{x\mathrm{d,R}}<0\\
    \end{array}
    \right..
    \label{hb22}
\end{equation}
As they pointed out, 
interpenetration of dust streams remains impossible even with this numerical flux 
because each cell cannot contain multiple velocities 
in the fluid approximation.
We call their method \texttt{HB22}.

\subsubsection{ Dust Fluxes Consistent with the Gas Transport,
\texttt{HLLgd}}\label{sec:HLLgd}

\citet{Verrier2025A&A...701A.174V} 
pointed out that dust solvers independent of the gas solver 
suffer from artificial variations in the dust--to--gas mass ratio in the strong-coupling limit because of inconsistencies between the dust and gas fluxes.
In the strong-coupling limit, $|\vvd-\vvg|$ is extremely small, 
and the gas and dust behave as a single fluid. 
In such a case, the mass transport of the dust fluid should be consistent with that of the gas fluid.
We call their method \texttt{HLLgd}.

In the HLL solver for the gas,  
two signal velocities, $S_\mathrm{g,L}$ and $S_\mathrm{g,R}$ ($S_\mathrm{g,L}<S_\mathrm{g,R}$), are defined. 
For the pressureless dust fluid, by contrast, the two characteristic velocities are 
$v_{x\mathrm{d,L}}$ and $v_{x\mathrm{d,R}}$. 

First, let us consider the case where the gas flux is determined by the intermediate state, 
$S_\mathrm{g,L}\le 0 \le S_\mathrm{g,R}$. 
\citet{Verrier2025A&A...701A.174V} adopt the dust flux
\begin{equation}
    \bm{F}_\mathrm{d}(S_\mathrm{g,L}\le 0\le S_\mathrm{g,R}) = \left\{
    \begin{array}{ll}
        \bm{F}_\mathrm{d,HLL} & S_\mathrm{g,L} < v_{x\mathrm{d,L}}, 
        v_{x\mathrm{d,R}} < S_\mathrm{g,R}\\
        \bm{F}_\mathrm{d,LLF} & \mathrm{otherwise.} 
    \end{array}
    \right.
    \label{Fdint}
\end{equation}
When the Riemann fan of the dust lies inside that of the gas, the HLL flux \citep{HartenLaxvanLeer1983} 
is also used for the dust: 
\begin{equation}
    \bm{F}_\mathrm{d,HLL} = 
    \frac{S_\mathrm{g,R}\bm{F}_\mathrm{d,L} - S_\mathrm{g,L}\bm{F}_\mathrm{d,R}
    + S_\mathrm{g,L}S_\mathrm{g,R} (\bm{U}_\mathrm{d,R} - \bm{U}_\mathrm{d,L})
    }{S_\mathrm{g,R}-S_\mathrm{g,L}},
    \label{hllgd}
\end{equation}
where $\bm{U}_\mathrm{d} =  (\rhod,\rhod\vvd)$ and
the signal velocities of the gas are used.
Otherwise, the local Lax--Friedrichs (LLF) flux for the dust is used:
\begin{equation}
    \bm{F}_\mathrm{d,LLF} = 
    \frac{\bm{F}_\mathrm{d,L} + \bm{F}_\mathrm{d,R}}{2}
    - \frac{S_\mathrm{LLF}}{2}
    (\bm{U}_\mathrm{d,R} - \bm{U}_\mathrm{d,L}),
\end{equation}
where $S_\mathrm{LLF} = \max(|v_{x\mathrm{d,L}}|, |v_{x\mathrm{d,R}}|)$.

Second, let us consider the case where the gas flux is determined by 
either the left or right state.
When the gas flux is determined by the left or right state, \texttt{HLLgd} uses the dust flux from 
the same state. Otherwise, the LLF flux is used.
\begin{equation}
    \bm{F}_\mathrm{d}(S_\mathrm{g,L}>0) = \left\{
    \begin{array}{ll}
        \bm{F}_\mathrm{d,L} & v_{x\mathrm{d,L}}>0\;\;\mathrm{and}\;\; v_{x\mathrm{d,R}} >0\\
        \bm{F}_\mathrm{d,LLF} & \mathrm{otherwise} 
    \end{array}
    \right.
    \label{Fdl}
\end{equation}
and 
\begin{equation}
    \bm{F}_\mathrm{d}(S_\mathrm{g,R}<0) = \left\{
    \begin{array}{ll}
        \bm{F}_\mathrm{d,R} & v_{x\mathrm{d,L}}<0\;\;\mathrm{and}\;\; v_{x\mathrm{d,R}} <0\\
        \bm{F}_\mathrm{d,LLF} & \mathrm{otherwise.} 
    \end{array}
    \right.
    \label{Fdr}
\end{equation}

\subsection{
Difficulties in Extending the \texttt{HLLgd} Approach to {HLLD}
}\label{sec:difficulty_hlld}

It is natural to consider extensions of \texttt{HLLgd} to less dissipative Riemann solvers,
such as HLLC and HLLD, because the HLL solver introduces relatively large numerical diffusion.
In this section, we discuss why straightforward extensions are not trivial.

It is useful to first examine why \texttt{HLLgd} works.
The applicability of \texttt{HLLgd} to both strong- and weak-coupling regimes 
relies on a generic property of the HLL solver.
The HLL flux requires only the left and right states and two signal velocities (see Equation (\ref{hllgd})), 
and does not require the detailed internal wave structure of the system.
\texttt{HLLgd} remains applicable in weak-coupling regimes because 
replacing the dust signal velocities with the gas signal velocities changes only the signal envelope.
This replacement may nevertheless introduce numerical diffusion associated with the gas signal speeds.

The situation is different for HLLD.
The HLLD flux is constructed from intermediate states associated with the internal wave structure of the MHD equations.
These intermediate states are determined by jump conditions involving gas pressure and Maxwell stress.
In a neutral pressureless dust fluid, the corresponding pressure and Maxwell stress are not defined.
Thus, there is no straightforward way to construct an HLLD dust flux whose 
mass component is consistent with the HLLD gas mass flux.
This presents a fundamental difficulty for a direct extension of the \texttt{HLLgd} construction to HLLD.

\subsection{A Split-Flux Method}\label{sec:newmethod}

Based on the arguments in Section \ref{sec:difficulty_hlld}, 
we do not construct a full HLLD-type Riemann solver for pressureless dust.
Instead, we propose a Split-Flux approach that is consistent with the HLLD gas solver 
in the strong-coupling regime while remaining applicable
in marginally and weakly coupled regimes.

Since the HLLD gas flux is associated with intermediate states determined by the gas dynamics, including 
gas pressure and Maxwell stress, the dust 
flux should satisfy the following three
requirements:
\begin{itemize}
    \item The dust flux should not be constructed from the 
    HLLD gas momentum flux, because the HLLD intermediate states involve gas pressure and Maxwell stress, which do not act on neutral pressureless dust.
    \item When the dust--gas drift is unresolved, the dust mass flux should be consistent with the HLLD gas mass flux.
    \item In weak-coupling regimes, the dust flux should reduce to an ordinary pressureless dust flux. 
    Otherwise, transport driven by the gas pressure and Maxwell stress would be 
    artificially incorporated into the dust dynamics.
\end{itemize}

Based on these requirements, the dust flux is constructed in the following 
two steps.
In Section \ref{sec:splitflux_stronglim},
we first construct a coupled-limit dust flux consistent with the HLLD gas mass flux.
We then blend this coupled-limit flux with the pressureless flux in Section \ref{sec:splitflux_blend}.

\subsubsection{Gas-Carried and Relative Dust Fluxes}\label{sec:splitflux_stronglim}

In strong-coupling regimes, the dust and gas move as a single fluid.
Our strategy differs from that of \texttt{HLLgd} in how the gas mass flux is used to construct the dust flux.
Instead of applying an HLLD-type flux 
directly to the dust solver, we start from an exact decomposition of the physical 
dust flux into two components: 
the component carried by the gas,
$\bm{F}_\mathrm{d,g}=\bm{U}_\mathrm{d} v_{x\mathrm{g}}$, and 
the component associated with the relative motion between the gas and dust,
$\bm{F}_\mathrm{d,rel}=\bm{U}_\mathrm{d}(v_{x\mathrm{d}}-v_{x\mathrm{g}})$. 
The dust flux is then written as 
\begin{equation}
    \bm{F}_\mathrm{d} = \bm{F}_\mathrm{d,g} + \bm{F}_\mathrm{d,rel}.
    \label{dustmassflux_div}
\end{equation}
The dust flux carried by the gas is determined from the HLLD gas mass
flux, $(F_\mathrm{g})_{\rhog}^\mathrm{HLLD}$, as
\begin{equation}
    \bm{F}_\mathrm{d,g} = \left\{
    \begin{array}{ll}
        (\bm{U}_\mathrm{d,L}/\rho_\mathrm{g,L})({F}_\mathrm{g})_{\rhog}^\mathrm{HLLD}
        & (F_\mathrm{g})^\mathrm{HLLD}_{\rhog}\ge 0
        \vspace{1mm} 
        \\ 
        (\bm{U}_\mathrm{d,R}/\rho_\mathrm{g,R})({F}_\mathrm{g})^\mathrm{HLLD}_{\rhog} & (F_\mathrm{g})^\mathrm{HLLD}_{\rhog}<0.\\ 
    \end{array}
    \right.
    \label{Fdg}
\end{equation}
This construction uses the gas mass flux provided by the HLLD solver 
without requiring the corresponding intermediate-state structure for the dust 
fluid. The same construction can in principle be applied to other gas solvers 
using their gas mass fluxes.
For the relative motion flux, we adopt the \texttt{HB22}-type solver, 
\begin{equation}
    \bm{F}_\mathrm{d,rel} = \left\{
    \begin{array}{ll}
         \bm{U}_\mathrm{d,L}\Delta {v}_{x\mathrm{L} }
         & \Delta v_{x\mathrm{L}} \ge 0,~\Delta v_{x\mathrm{R}}\ge 0\\
         \bm{U}_\mathrm{d,R}\Delta {v}_{x\mathrm{R}}
         & \Delta v_{x\mathrm{L}} \le 0,~\Delta v_{x\mathrm{R}}\le 0\\
         {0} & \Delta v_{x\mathrm{L}} < 0,~\Delta v_{x\mathrm{R}}>0\\
         \bm{U}_\mathrm{d,L}\Delta {v}_{x\mathrm{L}}
      + \bm{U}_\mathrm{d,R}\Delta {v}_{x\mathrm{R}}
         & \Delta v_{x\mathrm{L}} > 0,~\Delta v_{x\mathrm{R}}<0,\\
    \end{array}
    \right.
    \label{Fdrel}
\end{equation}
where $\Delta \bm{v} = \vvd - \vvg$.
It is worth noting that $({F}_\mathrm{d,rel})_{\rhod}$ is responsible 
for changing the dust--to--gas mass ratio $f_\mathrm{dg}$,
which satisfies 
\begin{equation}
    \frac{D_\mathrm{g} f_\mathrm{dg}}{Dt} = - \frac{1}{\rhog} \bm{\nabla} \cdot \left\{
     \rhod (\vvd - \vvg) 
    \right\},
    \label{epsilon_eq}
\end{equation}
where $D_\mathrm{g}/Dt = \partial /\partial t + \vvg \cdot\bm{\nabla}$.
The $x$ component of the relative dust mass flux term in Equation 
(\ref{epsilon_eq}) corresponds to the mass component of $\bm{F}_\mathrm{d,rel}$, the second 
term on the right-hand side of Equation (\ref{dustmassflux_div}).
When the dust and gas are tightly coupled at the grid scale,
Equation (\ref{epsilon_eq}) shows that $D_\mathrm{g}f_\mathrm{dg}/Dt \sim 0$,
indicating that $f_\mathrm{dg}$ is advected with the local gas velocity.
Accordingly, when the relative-motion component is small, 
the dust mass flux is dominated by the gas-carried component,
$(F_\mathrm{d,g})_{\rhod}$, and becomes consistent with the gas mass flux.

The dust flux obtained from Equations~(\ref{Fdg}) and (\ref{Fdrel}) is denoted by $(\bm{F}_\mathrm{d})_{\mathrm{cpl}}$. 
This flux is constructed from the HLLD gas mass flux and therefore
recovers gas-carried transport of both dust mass and 
dust momentum in the strong-coupling limit, without 
using the HLLD gas momentum flux.

\subsubsection{Blending Dust Mass and Momentum Fluxes}\label{sec:splitflux_blend}

As discussed in Section \ref{sec:introduction}, previous studies 
\citep{LaibePrice2012MNRAS.420.2345L,Verrier2025A&A...701A.174V}
suggest a resolution criterion of the form 
$\Delta x \lesssim c_\mathrm{s}\tstopd$.
Motivated by these criteria, 
we introduce the following dimensionless parameter $\taud$
\begin{equation}
    \taud = S_\mathrm{char}\frac{t_\mathrm{stop,d}}{\Delta x},
    \label{taud}
\end{equation}
where $S_\mathrm{char}$ is a characteristic speed specified in Section \ref{sec:chi_taud}.

The parameter $\taud$ can be interpreted as 
a dimensionless measure of whether the dust--gas drift is resolved 
on the grid.
It compares the characteristic stopping length,  
$S_\mathrm{char}t_\mathrm{stop,d}$, with the cell size.
For $\taud \gtrsim 1$, the dust--gas relative motion is resolved by the mesh, 
and the evolution of the dust density can be captured by 
the pressureless mass flux. 
By contrast, for $\taud\ll 1$, the relative motion between the dust and gas 
is unresolved. In this regime, inconsistencies between the 
dust mass flux and the gas mass flux can produce numerical errors in $f_\mathrm{dg}$.
Therefore, $(\bm{F}_\mathrm{d})_{\mathrm{cpl}}$ is appropriate, 
because the relative-motion component becomes small and the dust mass 
transport should follow the gas mass transport.

We blend the pressureless flux 
$(\bm{F}_\mathrm{d})_{\text{\texttt{HB22}}}$
and coupled-limit flux
$(\bm{F}_\mathrm{d})_{\mathrm{cpl}}$
using a transition function of $\tau_\mathrm{d}$ as follows:
\begin{equation}
    \bm{F}_\mathrm{d} =
    \chi(\tau_\mathrm{d}) 
    (\bm{F}_\mathrm{d})_{\mathrm{cpl}} 
    + \left\{ 1 - \chi(\tau_\mathrm{d}) \right\} 
    (\bm{F}_\mathrm{d})_{\text{\texttt{HB22}}},
    \label{splitflux}
\end{equation}
where $\chi$ is a smooth transition function that decreases from unity
for $\tau_\mathrm{d}\ll 1$ to zero for $\tau_\mathrm{d}\gg 1$.
The detailed choice of $\chi$ is discussed in Section \ref{sec:chi_taud}.

We refer to this formulation as the \texttt{Split-Flux} method. 
It decomposes the dust flux into gas-carried and relative-motion components 
and blends the resultant coupled-limit flux with the pressureless-fluid flux according to 
the degree to which the dust--gas drift is resolved.

\subsubsection{Dimensionless Dust Stopping Time $\taud$ and Transition Function $\chi$}\label{sec:chi_taud}

In the \texttt{Split-Flux} method, there is some flexibility 
in the choice of the characteristic speed $S_\mathrm{char}$ 
and in the form of the transition function $\chi$.

We first consider $S_\mathrm{char}$.
The dust--gas drift is often generated in shock fronts where the gas is suddenly decelerated and the dust is decelerated through drag
over a recoupling length $L_\mathrm{d}\sim c_\mathrm{s}\tstopd$
\citep{LovascioPaardekooper2019MNRAS.488.5290L}.
The resolution criterion based on 
$L_\mathrm{d}$ suggested by \citet{Verrier2025A&A...701A.174V} 
corresponds to $S_\mathrm{char}\sim c_\mathrm{s}$.

To extend this criterion to MHD, 
it is natural to adopt $S_\mathrm{char} = c_{\mathrm{f}x}$,
where $c_{\mathrm{f}x}$ is 
the fast-wave phase speed in the direction normal to the cell boundary.
When $c_{\mathrm{f}x}\tstopd/\Delta x < 1$, the recoupling length 
for a fast shock is not resolved.

Adopting $S_\mathrm{char}=c_{\mathrm{f}x}$ may overestimate 
the relevant stopping length because 
MHD has slow and Alfv\'en waves whose phase speeds are smaller 
than the fast-wave phase speed.
Especially in a strongly magnetized medium, 
the actual stopping length in sub-Alfv\'enic flows is expected to be
smaller than that predicted by $c_{\mathrm{f}x}\tstopd$.
In such a case, 
the pressureless flux may be used in regions 
where $c_\mathrm{s}\tstopd < \Delta x$ 
but $c_{\mathrm{f}x}\tstopd > \Delta x$.
We will discuss this issue in Section \ref{sec:sod}.

The use of a gas characteristic speed can also be motivated by the terminal-velocity estimate.
If the gas acceleration is produced by pressure or magnetic stresses varying over a length scale $\Delta x$, its magnitude is roughly
\begin{equation}
a_\mathrm{g} \sim \frac{c_\mathrm{f}^2}{\Delta x},
\end{equation}
where $c_\mathrm{f}$ denotes a representative fast-wave speed.
In the terminal-velocity regime, the dust--gas drift velocity is then estimated as
\begin{equation}
|\Delta \bm{v}| \sim a_\mathrm{g} t_\mathrm{stop,d}
\sim c_\mathrm{f}\left(\frac{c_\mathrm{f}t_\mathrm{stop,d}}{\Delta x}\right).
\end{equation}
Thus, $\taud=c_\mathrm{f}t_\mathrm{stop,d}/\Delta x$ corresponds to the expected drift velocity normalized by the gas characteristic speed,
\begin{equation}
\taud \sim \frac{|\Delta \bm{v}|}{c_\mathrm{f}},
\end{equation}
when the terminal-velocity estimate is applicable.
This interpretation supports the use of $\taud$ as a transition parameter. 
For $\taud\ll 1$, the drift velocity is expected to be small compared 
with the gas characteristic speed, whereas for $\taud\gtrsim 1$, 
the pressureless-fluid flux should be 
retained when the stopping length is resolved, even if 
the stopping time itself is short.

Adopting $S_\mathrm{char} = c_{\mathrm{f}x}$ may underestimate 
the stopping length associated with the dust--gas drift,
$|\dvv|\tstopd$, when $|\dvv|>c_{\mathrm{f}x}$.
We therefore adopt the following formula:
\begin{equation}
S_\mathrm{char} = \max\left( |\Delta \bm{v}|, c_{\mathrm{f}x}\right).
\label{Schar}
\end{equation}

We emphasize that 
the parameter $\taud$ is not a direct measure of 
the physical dust--gas coupling strength. Instead, $\taud$
evaluates whether the dust--gas drift is resolved on the grid.
Consequently, even in strongly coupled regimes, 
the pressureless dust flux is used when the dust--gas drift is sufficiently resolved.

The functional form of $\chi$ is also not unique.
The requirements are that $\chi\rightarrow 1$ for 
$\tau_\mathrm{d}\ll 1$,
$\chi\rightarrow 0$ for $\tau_\mathrm{d}\gg 1$, and that $\chi$ smoothly
transitions from 1 to 0 around $\tau_\mathrm{d}\sim 1$.
The former limit recovers the dust mass flux consistent with the gas mass flux, 
whereas the latter recovers the pressureless-fluid dust flux.
In this paper, we adopt
\begin{equation}
    \chi_{i+\frac{1}{2},j,k} = \exp\left( - \tau_{\mathrm{d},i+\frac{1}{2},j,k}^2\right)
    \label{transitionfunc}
\end{equation}
as a smooth transition function,
where $(i,j,k)$ are the coordinate indices of the cell centers,
and $x_{i+1/2}$ is the $x$ coordinate 
of the cell boundary between cells $i$ and $i+1$.
In Equation (\ref{transitionfunc}), 
$\tau_{\mathrm{d},i+{1}/{2},j,k}$ is the arithmetic average 
of $\tau_{\mathrm{d},i,j,k}$
and $\tau_{\mathrm{d},i+1,j,k}$.
As shown in Appendix \ref{app:chi}, the detailed functional 
form of $\chi$ has only a minor effect on the overall deviation
of $f_\mathrm{dg}$ from its initial value, 
although the localized deviations are moderately sensitive to 
the choice of $\chi$.


\begin{table}
  \caption{List of the solvers considered in this paper}
  \begin{center}
  \begin{tabular}{ccc}
      \hline
      & Gas Solver & Dust Solver   \\ 
      \hline
      \texttt{HB22}$^*$  & HLLD & Equation (\ref{hb22}) \\
      \texttt{HLLgd}$^\dag$ & HLLE & Equations (\ref{Fdint}), (\ref{Fdl}), and (\ref{Fdr})  \\
     \texttt{Split-Flux} & HLLD & Equation (\ref{splitflux})\\
      \hline
    \end{tabular}
    \label{tab:method}
  \end{center}
$^*$ \citet{HuangBai2022ApJS..262...11H}  \\ 
$^\dag$ \citet{Verrier2025A&A...701A.174V}
\label{tab:method}
\end{table}

\section{Numerical Experiments} \label{sec:experiments}

In the numerical experiments presented in this section, 
we restrict the nonlinear tests to small dust--to--gas mass ratios, whereas 
the linear dusty MHD wave tests adopt $f_\mathrm{dg}\sim 1$.
In the nonlinear tests, 
the gas dynamics is only weakly affected by the dust.
This choice isolates the dust-transport errors caused by inconsistencies
between the dust and gas numerical fluxes. 
When $f_\mathrm{dg}\sim 1$, 
drag-induced energy dissipation and total energy conservation
can become important in nonlinear problems. 
A detailed investigation of these issues is beyond the scope of this study.

\begin{figure*}[htpb]
    \centering
    \begin{tabular}{cc}
    \includegraphics[width=0.4\linewidth]{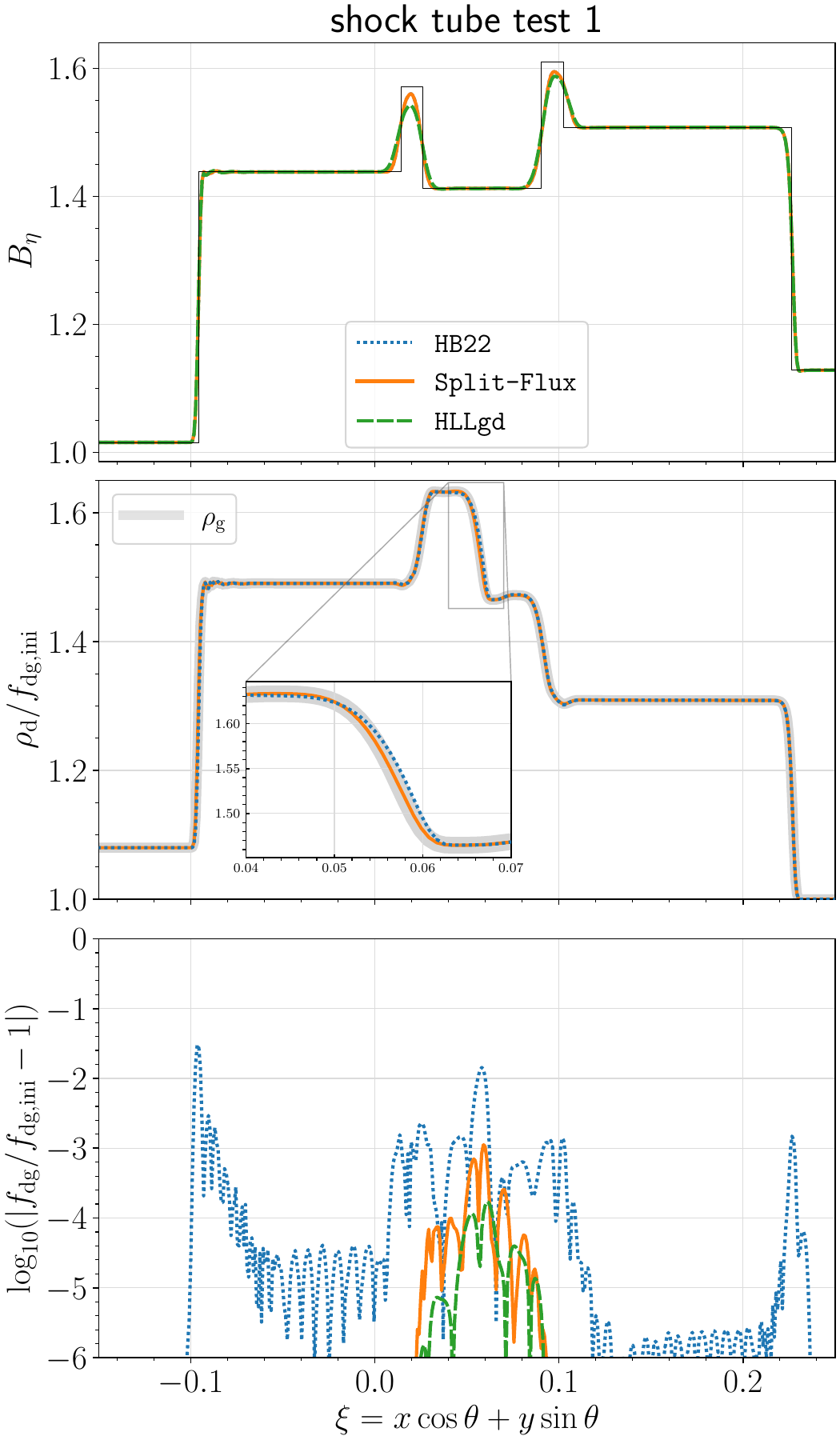}
    &
    \includegraphics[width=0.4\linewidth]{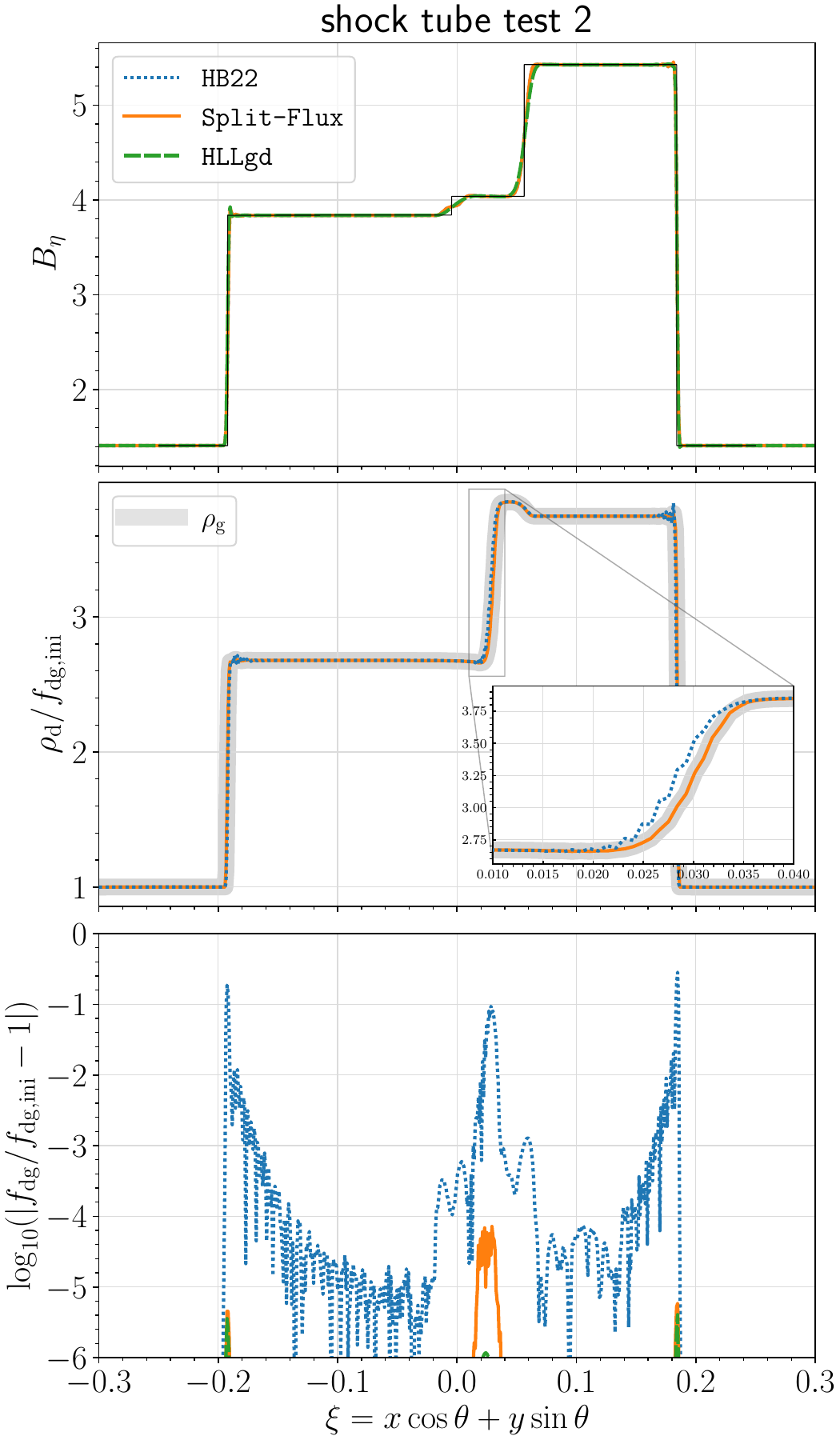}\\
    \end{tabular}
    \caption{
    Results of the two shock tube tests for
    \texttt{HB22}, \texttt{Split-Flux}, and \texttt{HLLgd}.
    The left and right columns correspond to the first and second shock tube tests,
    respectively.
    The top panels show the tangential magnetic field $B_\eta$,
    the middle panels show the dust density, $\rhod/f_\mathrm{dg,ini}$, and the bottom panels show $\log_{10}|f_\mathrm{dg}/f_\mathrm{dg,ini}-1|$,
where $f_\mathrm{dg,ini}$ is the initial dust--to--gas mass ratio.
    In the middle panels, the thick gray lines correspond to the gas density, and the results of \texttt{HLLgd} 
    are not displayed.
    }
    \label{fig:shocktube}
\end{figure*}

\subsection{Two-dimensional Shock Tube Problems}\label{sec:shock}
To examine the performance of the \texttt{Split-Flux} method 
in the strong-coupling limit, 
we conduct two-dimensional shock tube tests with $\tstopd=10^{-9}$.
The normal to the initial discontinuity is inclined by 
an angle $\theta$ with respect to the $x$-axis, where $\tan\theta=2$.
The $x$ range of the simulation box is set to $-1\le x\le 1$.
The $y$ range is chosen so that the cells are square.
We use $N_x=1024$ and $N_y=8$ cells.
Outflow boundary conditions are imposed at the $x$ boundaries, 
and shifted outflow boundary conditions are imposed at the $y$ boundaries \citep{Mignone2010JCoPh.229.2117M}.
The initial condition is specified in the rotated coordinate system 
$(\xi,\eta,z) = (x\cos\theta+y\sin\theta,-x\sin\theta + y\cos\theta,z)$.
The initial dust--to--gas mass ratio is $10^{-3}$, and the dust fluid initially 
moves at the gas velocity.

The first shock tube test contains all types of MHD discontinuities \citep{DaiWoodward1994JCoPh.111..354D}.
The initial condition is $(\rho_\mathrm{g},P_\mathrm{g},v_{\mathrm{g}\xi},v_{\mathrm{g}\eta},v_{\mathrm{g}z},B_{\eta},B_{z})
= (1.08,~0.95,~1.2,~0.01,~0.5,~3.6/\sqrt{4\pi},~2/\sqrt{4\pi})$ for $\xi<0$ and 
$(1,~1,~0,~0,~0,~4/\sqrt{4\pi},~2/\sqrt{4\pi})$ for $\xi>0$ with $B_\xi = 4/\sqrt{4\pi}$.
The second shock tube test involves strong fast shocks \citep{DaiWoodward1994JCoPh.111..354D}.
The initial condition is $(\rho_\mathrm{g},P_\mathrm{g},v_{\mathrm{g}\xi},v_{\mathrm{g}\eta},v_{\mathrm{g}z},B_{\eta},B_{z})
= (1,~20,~-10,~0,~0,~5/\sqrt{4\pi},~0)$ for $\xi<0$ and 
$(1,~1,~10,~0,~0,~5/\sqrt{4\pi},~0)$ for $\xi>0$ with $B_\xi = 5/\sqrt{4\pi}$.
For both shock tube tests, we adopt an adiabatic index of $\gamma=5/3$.

The top panels of Figure \ref{fig:shocktube} show the tangential magnetic field $B_\eta$.
Since $f_\mathrm{dg}$ is extremely small, the gas dynamics is not 
affected by the dust.
Therefore, differences in $B_\eta$ primarily reflect the choice of the gas solver.
The \texttt{HB22} and \texttt{Split-Flux} methods 
show almost identical results because both use HLLD as the gas solver (Table \ref{tab:method}).
Differences between the HLLD-based methods
(\texttt{HB22} and \texttt{Split-Flux}) 
and \texttt{HLLgd} are clearly 
seen in $B_\eta$ for the first shock tube test.
The HLLD-based methods capture rotational discontinuities and slow shocks more sharply than \texttt{HLLgd} \citep{MiyoshiKusano2005JCoPh.208..315M}.

The dust mass densities are compared with the gas mass densities in 
the middle panels of Figure \ref{fig:shocktube}.
Since $\tstopd$ is extremely small, $\rhod/f_\mathrm{dg,ini}$  
should closely follow $\rhog$ throughout the evolution.
The magnified views near the contact discontinuities show that 
\texttt{HB22} produces an artificial 
offset between the dust and gas contact discontinuities, whereas the $\rhod/f_\mathrm{dg,ini}$ 
profiles coincide with 
the $\rhog$ profiles for \texttt{Split-Flux}.

The bottom panels of Figure \ref{fig:shocktube}, which show the deviation of $f_\mathrm{dg}$ 
from its initial value, more clearly reveal the mismatch between $\rhod$ and $\rhog$.
For both shock tube tests, when \texttt{HB22} is used in the strong-coupling limit,
relatively large errors appear not only near the contact discontinuities 
but also near the other discontinuities.
By contrast, 
\texttt{Split-Flux} keeps the errors in $f_\mathrm{dg}$ as small as those obtained with \texttt{HLLgd},
while retaining the sharper gas dynamics obtained with the HLLD gas solver.

The errors in $f_\mathrm{dg}$ are concentrated near the discontinuities.
In smoothly varying regions, \texttt{HB22} also produces relatively small $f_\mathrm{dg}$ errors.
This behavior can be understood by the fact that the coupled-limit and 
the pressureless fluxes are expected to differ most strongly across sharp structures, whereas 
both fluxes provide similar mass fluxes in smoothly varying regions. 
Thus, inconsistencies in the dust--to--gas mass ratio tend to 
be generated primarily around sharp structures, such as shocks and contact discontinuities.

\subsection{Sod Solution with a Parallel Magnetic Field}\label{sec:sod}

As discussed in Section \ref{sec:chi_taud},
the use of the fast-wave speed in $\taud$ 
may overestimate the stopping length 
when the gas motion is sub-Alfv\'enic 
in a strongly magnetized medium.
In such cases, 
the pressureless flux can be used when the actual 
dust--gas drift is not well resolved.
To examine this possibility, we consider a Sod shock tube propagating parallel to a strong magnetic field because 
the gas motions are often guided along magnetic field lines. 
We adopt an adiabatic index of $\gamma=1.4$.

As in the shock tube problems presented in Section \ref{sec:shock}, 
the initial discontinuity is tilted by the angle $\theta$ 
relative to the x-axis.
The $x$ range of the simulation box is set to $-1\le x\le 1$.
The $y$ range is chosen so that the cells are square.
We use $N_x=512$ and $N_y=8$ cells.
The initial dust--to--gas mass ratio is $10^{-3}$, and 
the dust fluid is initially static.

We compare a non-magnetized case, $B_\xi=0$, with a strongly 
magnetized case, $B_\xi=30$.
The plasma betas for $B_\xi=30$ in the left and right states are 
$2\times 10^{-3}$ and $2\times 10^{-4}$, respectively.
Because the magnetic field is parallel to the propagation direction, 
the corresponding solution is identical to the hydrodynamic Sod solution.

\begin{figure}[htpb]
    \centering
    \includegraphics[width=0.9\linewidth]{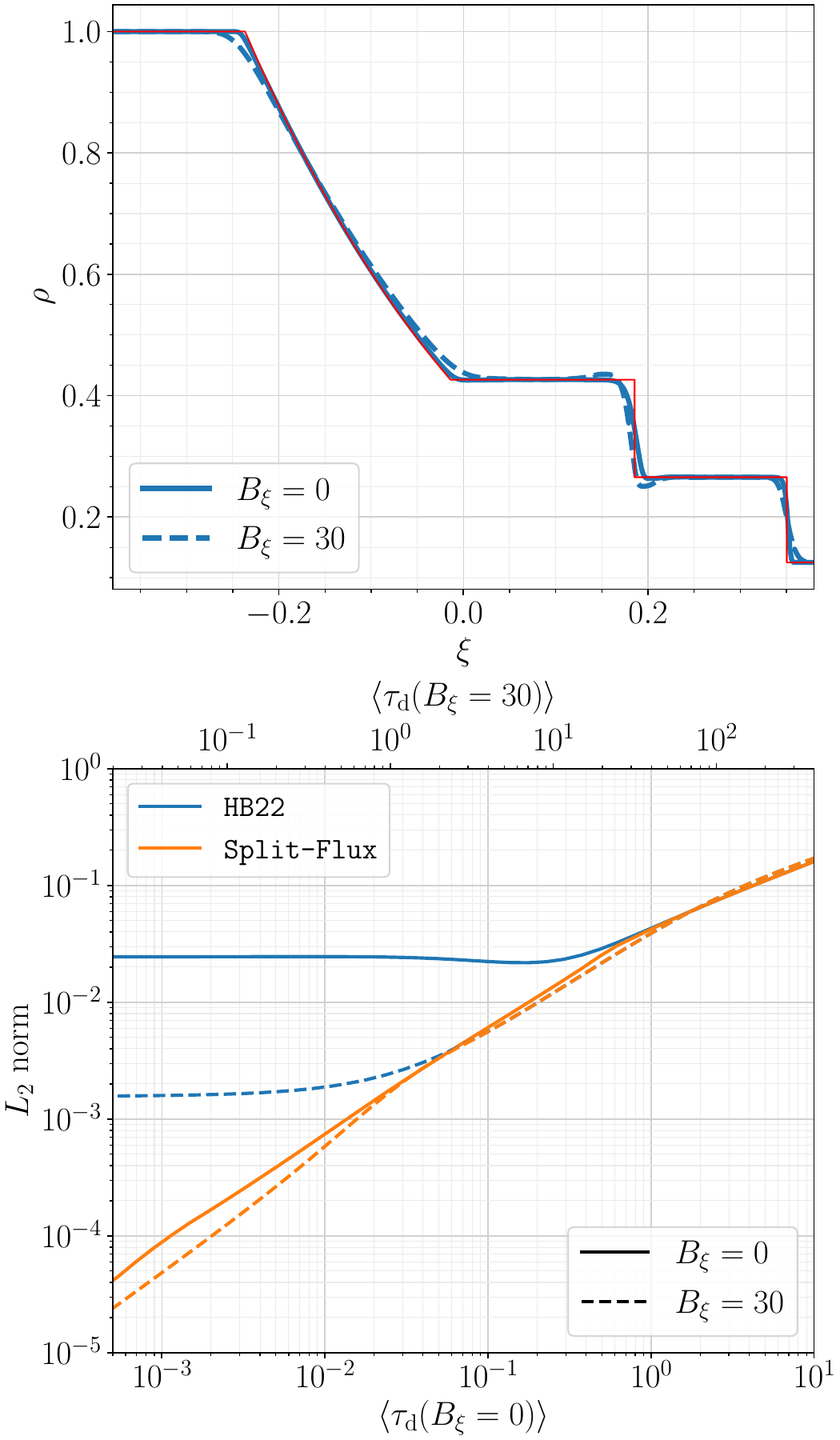}
    \caption{
    (Top panel) 
    Gas density profiles for $B_\xi=0$ (solid) and 
    $B_\xi=30$ (dashed) at $t=0.2$, obtained with \texttt{HB22}.
    The red lines correspond to the exact solution.
    (Bottom panel) 
    The $L_2$ norm of $f_\mathrm{dg}/f_\mathrm{dg,ini}-1$
    for $B_\xi=0$ (solid) and $B_\xi=30$ (dashed).
    The blue and orange lines show the results obtained with \texttt{HB22} and 
    \texttt{Split-Flux}, respectively.
    The $L_2$ norms are computed over $|\xi|\le 0.38$, which contains the complete Riemann fan.
    For reference, the lower and upper horizontal axes show the volume-averaged values of 
    $\tau_\mathrm{d}$ with $c_\mathrm{f}=\sqrt{(\gamma \Pg + B_\xi^2)/\rhog}$ for
    $B_\xi=0$ and $B_\xi=30$, respectively.
    }
    \label{fig:sod}
\end{figure}

We first examine the dependence of the gas solution on $B_\xi$. 
The top panel of Figure \ref{fig:sod} shows that 
the density profile in the strongly magnetized case is more 
diffuse than that in the non-magnetized case.
This is because the MHD variables are updated using the
HLLD numerical fluxes along the $x$- and $y$-axes, which 
are not parallel to the magnetic field.
Consequently, the numerical dissipation depends on 
the field strength, producing more diffuse gas structures.

We next examine how $\rho_\mathrm{d}/f_\mathrm{dg,ini}$ approaches 
$\rhog$ as $t_\mathrm{stop,d}$ decreases.  
We compute the $L_2$ norms of $f_\mathrm{dg}/f_\mathrm{dg,ini}-1$
over $|\xi|\le 0.38$ at $t=0.2$.
The lower and upper horizontal axes in the bottom panel of Figure \ref{fig:sod} 
show the volume-averaged reference values of $\tau_\mathrm{d}(B_\xi) = t_\mathrm{stop,d}
\sqrt{(\gamma \Pg+ B_\xi^2)/\rhog} /\Delta x$, denoted by $\langle \taud\rangle$.
This reference quantity provides a volume-averaged estimate of $\taud$, whereas
the \texttt{Split-Flux} method uses the local face-centered value.

For both magnetic field strengths, 
the $L_2$ norms obtained with \texttt{HB22} and \texttt{Split-Flux} begin to 
separate around $\langle \taud(B_\xi)\rangle \sim 1$.
For $\langle \taud(B_\xi)\rangle \gtrsim 1$, 
the two methods give nearly identical $L_2$ norms.
As $\langle \taud(B_\xi)\rangle$ decreases below unity,  
the \texttt{HB22} norm eventually approaches a floor,
whereas 
the \texttt{Split-Flux} norm continues to decrease with 
decreasing $t_\mathrm{stop,d}$. 

Although the precise location of the transition may depend on the details of
the numerical schemes, the results suggest that $\taud$ based on the fast-wave speed 
provides a practical estimate of the transition between the coupled-limit and 
the pressureless fluxes.

For $B_\xi=30$, the larger fast-wave speed shifts this separation between \texttt{HB22} and 
\texttt{Split-Flux} toward a smaller $\tstopd$.
Nevertheless, 
the bottom panel of Figure \ref{fig:sod} shows that 
both $L_2$ norms for \texttt{HB22} and \texttt{Split-Flux} continue to decrease until 
$\langle \taud(B_\xi=30)\rangle$ reaches unity.
We attribute this behavior to the broader gas structures produced by numerical diffusion, which 
allow the dust to follow the gas dynamics over a larger number of cells.

This result suggests that the fast-wave speed 
provides a practical and conservative choice for $S_\mathrm{char}$,
even for 
sub-Alfv\'enic gas motions in a strongly magnetized medium.

\subsection{Dusty MHD Waves}\label{sec:mhddustywave}

We present an MHD extension of the
\textsc{dustywave} test, which is a linear wave test for dust--gas mixtures with drag 
\citep{LaibePrice2011MNRAS.418.1491L}.

Following \citet{Mignone2010JCoPh.229.2117M}, 
we consider linear waves propagating in an oblique direction.
The unperturbed state consists of static, uniform gas and dust 
with $\rhog=1$, $\Pg = 1$, and $\rhod=1$, embedded in a uniform 
magnetic field of $\bm{B}_0 = B_0(\cos (\pi/4), \sin(\pi/4), 0)$.
Three-dimensional simulations are performed.
The simulation domain is $0\le x\le 1$ and 
$0\le y,z\le 0.5$, with periodic boundary conditions in all directions.
The wavenumber vector is set to $\bm{k} = 2\pi(1,2,2)$.
We adopt an adiabatic index of $\gamma=5/3$.

\subsubsection{Uniform Grids}

We first consider uniform grids.
We examine both fast and slow waves over a parameter space spanned by ${B}_0$ and $\tstopd$.
For the magnetic field strength, we consider weakly and strongly magnetized 
cases, $B_0=0.1$ and $10$,
because these cases produce a large separation between the phase speeds of fast and slow waves.

For each parameter set, we conduct the simulations with $N_x=8$, $16$, $32$, $64$, $128$, and $256$.
The numbers of cells along the $y$- and $z$-axes are $N_y = N_x/2$ 
and $N_z = N_x/2$, 
respectively.
The dimensionless parameter $t_\mathrm{stop,d}c_\mathrm{ph}k$ characterizes 
the physical dust--gas coupling, where $c_\mathrm{ph}$ is the phase speed in the dust-free 
limit ($\rhod=0$), and $k \equiv |\bm{k}|$ is the wavenumber. The coupling becomes weak when $t_\mathrm{stop,d}c_\mathrm{ph}k\gtrsim 1$.
We consider two values of this parameter: $t_\mathrm{stop,d}c_\mathrm{ph}k=10^{-2}$ (strong coupling) 
and $1$ (marginal coupling). 
For each parameter set, the eigenfunctions are derived and then used 
as the initial conditions.
The amplitude is normalized so that the largest amplitude among all physical quantities is $10^{-6}$.

We compute the $L_1$ norm for the dust density perturbation as 
\begin{equation}
L_1 = \frac{1}{N_xN_yN_z}\sum_{i,j,k} |(\rho_{\mathrm{d}})_{i,j,k} 
- \rho_\mathrm{d,ana}(x_i,y_j,z_k,2\pi/\omega)|,
\end{equation}
where $(i,j,k)$ are the coordinate indices and $\rho_\mathrm{d,ana}(x,y,z,t)$ is the exact solution.
The $L_1$ norms are computed using the results after one wave period,
at $t=2\pi/\omega$, where $\omega$ is the wave frequency.

\begin{figure}[htpb]
    \centering
    \includegraphics[width=1.0\linewidth]{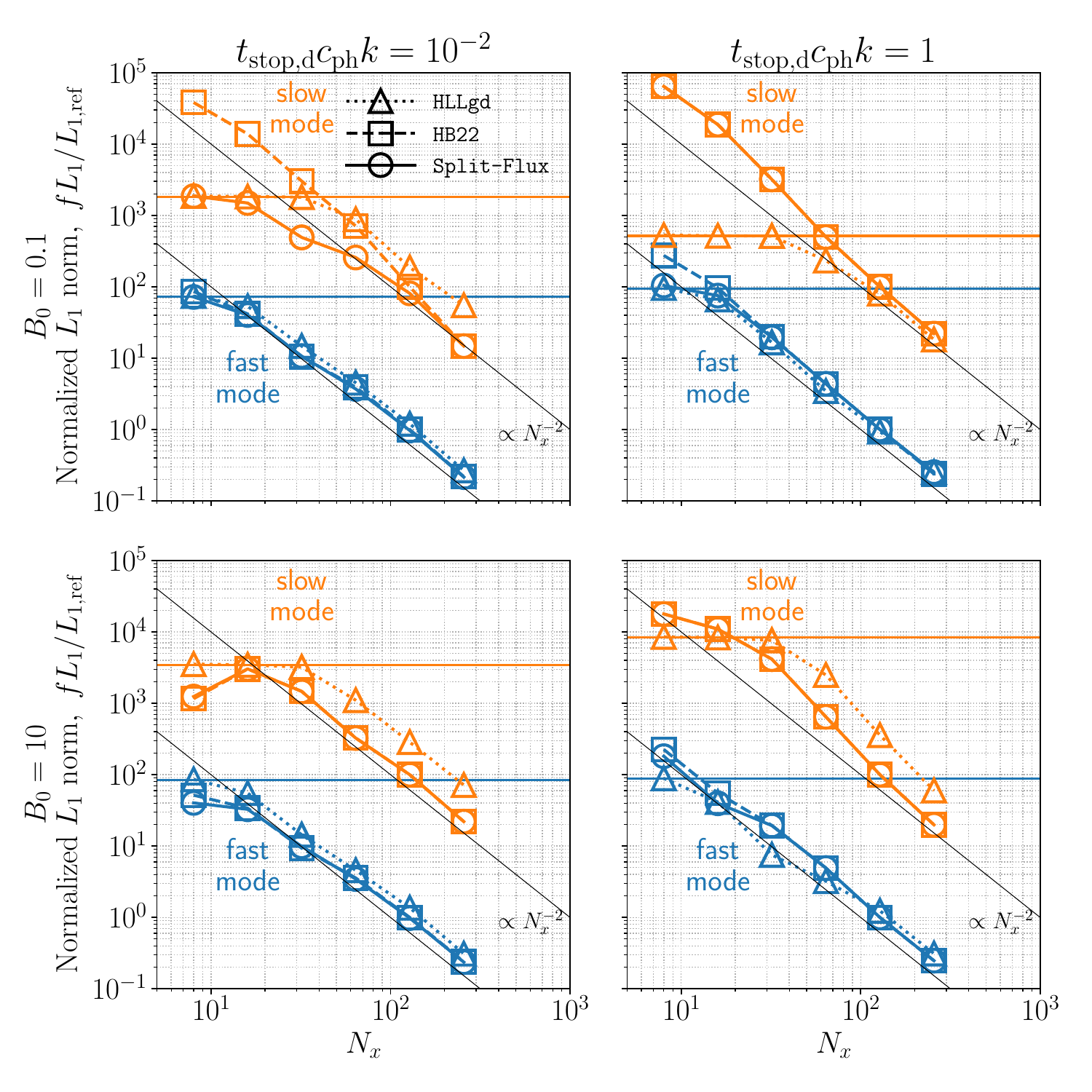}
    \caption{
    Convergence tests of MHD dusty fast waves (blue) and slow waves (orange)
    for \texttt{HLLgd} (dotted), \texttt{HB22} (dashed), and \texttt{Split-Flux} (solid).
    The $L_1$ norms of the dust density perturbations are shown 
    as a function of $N_x$.
    The left and right panels show the results with 
    $t_\mathrm{stop,d}c_\mathrm{ph}k=10^{-2}$ and $1$, respectively. 
    The upper and lower panels show the results with $B_0=0.1$ and $10$, respectively.
    For each parameter set $(t_\mathrm{stop,d}c_\mathrm{ph}k, B_0)$ and each mode,
    the horizontal line corresponds to the expected $L_1$ norm
    when the dust density perturbation is completely damped.
    The vertical axes show the normalized errors, 
     $fL_1/L_\mathrm{1,ref}$, where 
     the reference error $L_{1,\mathrm{ref}}$ is $L_1(N_x=128)$ for \texttt{HB22} 
     for each parameter set of
     ($t_\mathrm{stop,d}c_\mathrm{ph}k$, $B_0$).
     The factor $f$ is unity for the fast waves and 100 for the slow waves.
     In each panel, two solid straight lines show $L_1\propto N_x^{-2}$.
    }
    \label{fig:mhdwave_conv}
\end{figure}

Figure \ref{fig:mhdwave_conv} shows the results of the convergence tests on uniform grids. 
The vertical axes show the normalized error, $fL_1/L_{1,\mathrm{ref}}$, 
where $L_\mathrm{1,ref}$ is $L_1(N_x=128)$ for \texttt{HB22} and the scaling factor $f$ is introduced only to improve the visibility 
of the fast- and slow-wave errors in the same panels.

The horizontal lines indicate the normalized $L_1$ norms obtained 
if the dust density perturbation is completely damped and only the uniform background dust density remains.
The $L_1$ norms obtained with \texttt{HB22} can exceed 
the horizontal lines. 
Especially for the slow waves with $B_0=0.1$, the $L_1$ norms are
well above the horizontal lines.
This indicates that 
the residual dust density errors exceed those expected 
from complete damping of the wave amplitude.
Such dust density errors are associated with the pressureless nature 
of the dust fluid, which provides no physical mechanism 
for smoothing short-wavelength dust density structures.

The \texttt{HLLgd} results show that 
the $L_1$ norms approach the horizontal lines for all cases 
and the dust density perturbations are damped as $N_x$ decreases.
This is attributed to numerical diffusion introduced by the HLL-type 
flux, whose signal speeds are determined by the gas dynamics.
This numerical diffusion suppresses the spurious density structures 
associated with 
the pressureless nature, but also increases the dissipation of 
the physical dust density perturbations, especially for the slow modes.

The \texttt{Split-Flux} method 
achieves the smallest $L_1$ errors among the three methods in
the strongly coupled case, $t_\mathrm{stop,d}c_\mathrm{ph}k=10^{-2}$, 
particularly for the slow mode.
As $N_x$ increases, the $L_1$ norms for \texttt{Split-Flux} approach those 
for \texttt{HB22}. When $N_x\ge 128$, both are almost identical.
This behavior reflects the fact that the dust flux transitions from the coupled-limit flux to 
the pressureless flux.
Combining $t_\mathrm{stop,d}c_\mathrm{ph}k=10^{-2}$ and $c_\mathrm{f}\tstopd/\Delta x \sim 1$, 
we can define a critical $N_x$ for the slow wave with $B_0=0.1$ above which 
\begin{equation}
   N_{x,\mathrm{crit}} \sim 10^2 
   \left(\frac{c_\mathrm{ph}/c_\mathrm{f}}{0.055} \right)
   \left(\frac{kL_x}{6\pi}\right)
   \left(\frac{\tstopd c_\mathrm{ph} k }{10^{-2}}\right)^{-1},
   \label{Ncri}
\end{equation}
where $L_x$ is the box size along the $x$-axis.
The top-left panel of Figure \ref{fig:mhdwave_conv} shows that 
the transition occurs around $N_x \sim N_{x,\mathrm{crit}}$.

For $t_\mathrm{stop,d}c_\mathrm{ph}k = 1$, 
the $L_1$ norms for \texttt{Split-Flux} are nearly identical to those for \texttt{HB22}, 
as expected from the design of the transition function $\chi$ 
in the marginal- and weak-coupling regimes (Equation (\ref{splitflux})).

Consequently, for $t_\mathrm{stop,d}c_\mathrm{ph}k =1$,
the $L_1$ norms obtained with \texttt{Split-Flux} 
can exceed the complete-damping level, as do those of \texttt{HB22}.
This behavior reflects the intended transition to the pressureless solver rather 
than a numerical artifact specific to \texttt{Split-Flux}.
Reducing these errors arising from the pressureless nature 
is beyond the scope of this paper.

\subsubsection{Non-uniform Grids with Static Mesh Refinement}\label{sec:smr}

We next examine the three methods on grids with static mesh refinement.
In this test, $N_x$ corresponds to the number of cells along the $x$-axis 
at the base level.
The region defined by $0.25\le x\le 0.75$ and $0.125\le y,z\le 0.375$ is refined by one level with a refinement ratio of two.

We focus on the fast waves with $B_0=1$, for which all three methods provide comparable errors on uniform grids.
As in the uniform-grid tests, 
we adopt $t_\mathrm{stop,d}c_\mathrm{ph}k = 10^{-2}$ and $1$.
In this section, we show the results after one wave period.

We perform the SMR tests with $N_x=32$, $64$, $128$, and $256$.
For the fast wave with $B_0=1$, $c_\mathrm{ph}/c_\mathrm{f}\sim 1$.
The number of cells along the $x$-direction at which $\taud = 1$ at level $\ell$ 
is 
\begin{equation}
    N_{x,\mathrm{crit}}^{(\ell)}
    \sim 
    \frac{2\times 10^3}{2^{\ell}} \left(\frac{kL_x}{6\pi}\right)
     \left(\frac{\tstopd c_\mathrm{ph}k}{10^{-2}}\right)^{-1},
     \label{Ncr_smr}
\end{equation}
where $\ell=0$ and 1 denote the base and refined levels, respectively. 
For $t_{\mathrm{stop,d}}c_{\rm ph}k=10^{-2}$,
$N_{x,\mathrm{crit}}^{(0)}\sim2\times10^3$ 
and $N_{x,\mathrm{crit}}^{(1)}\sim 10^3$, 
both of which exceed the values of $N_x$ considered here.
The coupled-limit flux therefore remains dominant on both levels. 
For $t_{\mathrm{stop,d}}c_{\rm ph}k=1$, $N_{x,\mathrm{crit}}^{(0)}\sim20$ 
and $N_{x,\mathrm{crit}}^{(1)}\sim 10$. Thus, for $N_x\ge32$, the \texttt{HB22} flux dominates on both sides of the refinement boundaries.

\begin{figure}[htpb]
    \centering
    \includegraphics[width=1.0\linewidth]{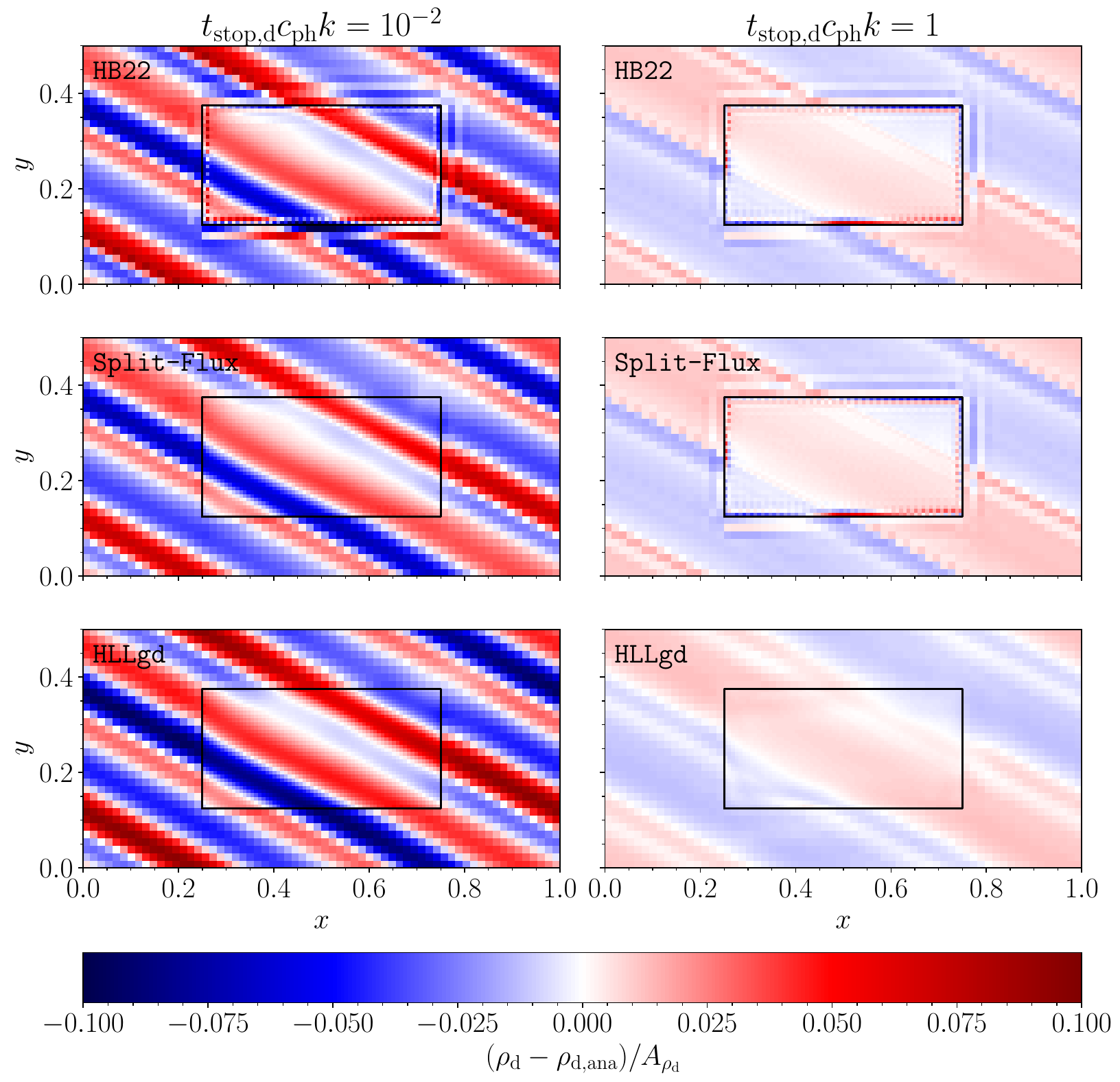}
    \caption{
    Deviation of $\rhod$ from the exact solution $\rho_\mathrm{d,ana}$ at the $z=0.26$ plane
    for \texttt{HB22} (top), \texttt{Split-Flux} (middle), and \texttt{HLLgd} (bottom).
    The results with $N_x=64$ are shown.
    The fast waves with $B_0=1$ are considered.
    The left and right panels correspond to the results with $t_\mathrm{stop,d}c_\mathrm{ph}k=10^{-2}$ and $1$, respectively.
    The plotted difference is normalized by the amplitude of the dust density perturbation, $A_{\rhod}$.
    The refined regions are enclosed by the rectangles.
    }
    \label{fig:mhdwave_smr_cmap}
\end{figure}

Figure \ref{fig:mhdwave_smr_cmap} shows the color maps 
of the difference between 
the dust density and the exact solution. 
We first consider the results with $t_\mathrm{stop,d}c_\mathrm{ph}k=10^{-2}$, 
which corresponds to the strong-coupling regime.
Prominent artificial structures with wavelengths close to the Nyquist wavelength 
are seen near the refinement boundaries in \texttt{HB22}.
For the other two methods, \texttt{Split-Flux} and \texttt{HLLgd}, 
no clear artificial structures are seen near the refinement boundaries.

Next, we investigate the marginal coupling case, $\tstopd c_\mathrm{ph}k=1$.
Figure \ref{fig:mhdwave_smr_cmap} shows that 
prominent artificial structures are also seen near the refinement boundaries in \texttt{HB22}.
Unlike in the strong-coupling case $\tstopd c_\mathrm{ph}k=10^{-2}$,
similar structures are also present in \texttt{Split-Flux} because 
$(\bm{F}_\mathrm{d})_{\mathtt{HB22}}$ contributes to 
the dust flux in \texttt{Split-Flux} (Equation (\ref{splitflux})).
In \texttt{HLLgd}, such artificial structures are not clearly seen 
near the refinement boundaries.

\begin{figure}[htpb]
    \centering
    \includegraphics[width=1.0\linewidth]{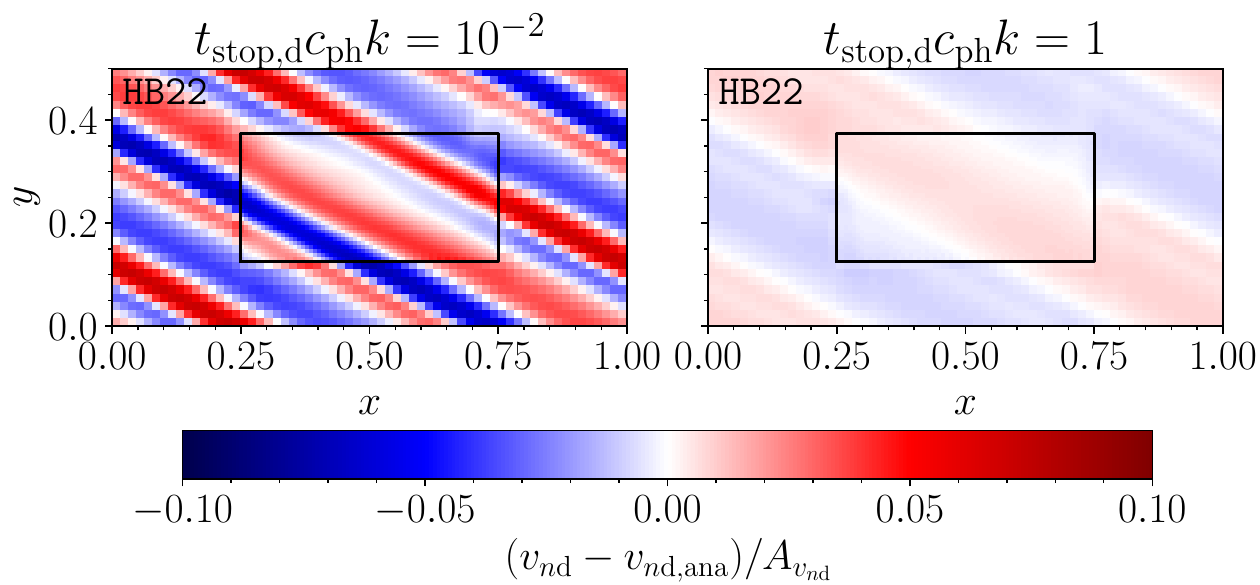}
    \caption{
    The same as Figure \ref{fig:mhdwave_smr_cmap} but for the dust velocity perturbations along $\bm{k}$,
    where $v_{n\mathrm{d}} \equiv \bm{v}_\mathrm{d}\cdot\bm{k}/|\bm{k}|$ is the dust velocity along 
    the wavenumber vector, $v_{n\mathrm{d},\mathrm{ana}}$ is the corresponding exact solution, and 
    $A_{v_{n\mathrm{d}}}$ is the amplitude of $v_{n\mathrm{d},\mathrm{ana}}$.
    Only the results of \texttt{HB22} with $t_\mathrm{stop,d}c_\mathrm{ph}k=10^{-2}$ and $1$ are shown.
    }
    \label{fig:mhdwave_smr_cmap_vn}
\end{figure}

Figure \ref{fig:mhdwave_smr_cmap_vn} shows that 
the dust velocity perturbations along $\bm{k}$ do not exhibit clear artificial structures near the refinement boundaries,
even for \texttt{HB22}.
We confirmed that the gas density and velocity fields are also free from such artificial structures.
As discussed in \citet{Verrier2025A&A...701A.174V}, artificial disturbances in $\rhod$ generated near the refinement boundary 
are not efficiently damped because 
the pressureless dust fluid provides no restoring force for density perturbations.
By contrast, as shown in Figure \ref{fig:mhdwave_smr_cmap_vn}, grid-scale disturbances in the dust velocity can 
be suppressed by drag from the gas.

In this test, convergence is measured using two error norms for 
the dust density, the $L_1$ and $L_\infty$ norms.
\begin{equation}
L_1 = \frac{\sum_m \Delta x_m^3|e_m|}{\sum_m \Delta x_m^3},\;\;\;
L_\infty = \max_m |e_m|,
\end{equation}
where $e_m = (\rhod)_m - \rho_\mathrm{d,ana}(x_m,y_m,z_m,t=2\pi/\omega)$, 
the index $m$ spans all the cells, and $\Delta x_m$ is the width 
of cell $m$.
The $L_1$ norm measures the overall error, 
whereas the $L_\infty$ norm is sensitive to  
localized numerical artifacts that can emerge around the mesh-refinement boundaries.

\begin{figure}[htpb]
    \centering
    \includegraphics[width=0.8\linewidth]{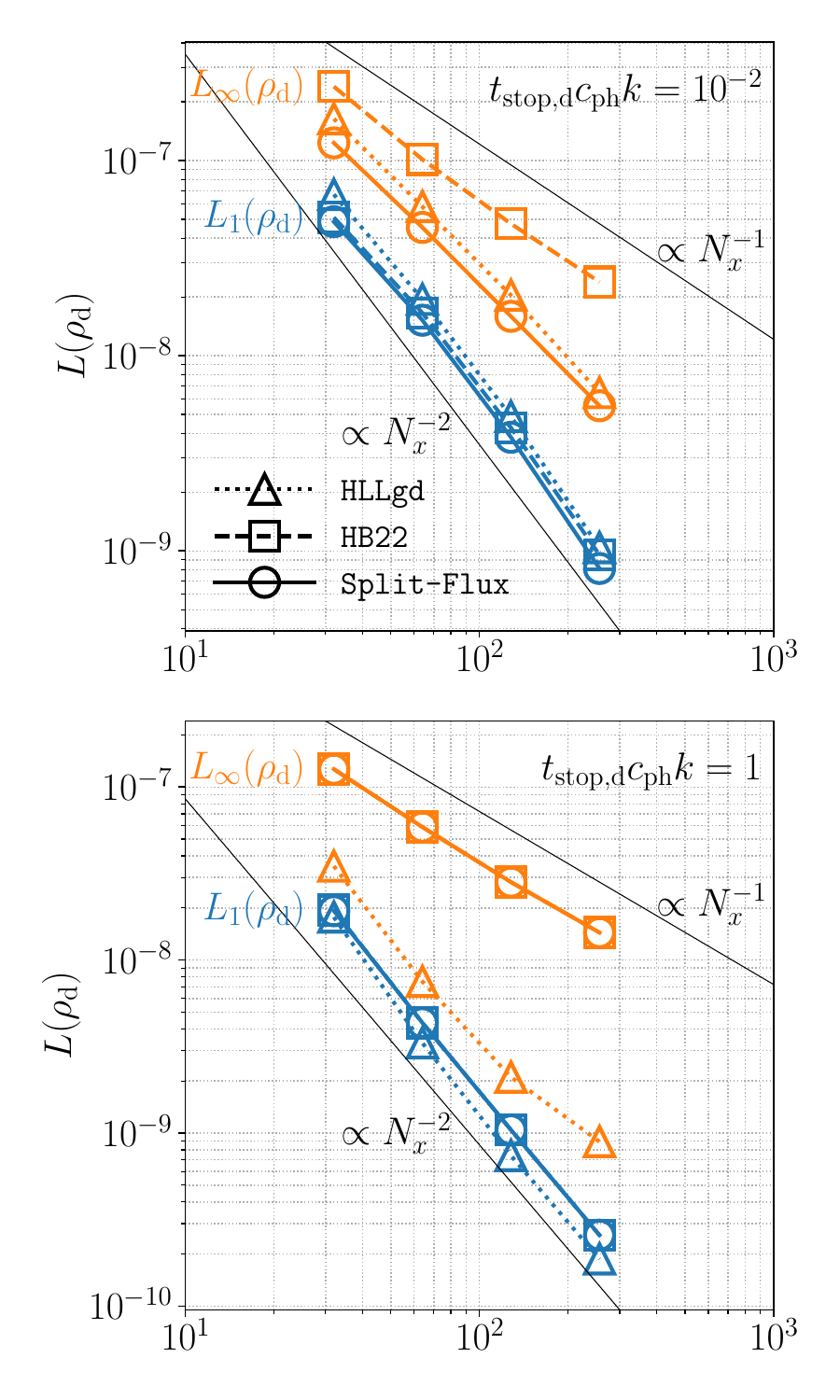}
    \caption{
     Convergence tests for fast waves with $t_\mathrm{stop,d}c_\mathrm{ph}k=10^{-2}$ (top) and 
     $1$ (bottom), obtained using \texttt{HLLgd}, \texttt{HB22}, and \texttt{Split-Flux}.
    The blue and orange lines show the $L_1$ and $L_\infty$ norms of the dust density errors, respectively.
    }
    \label{fig:mhdwave_smr_conv}
\end{figure}

Figure \ref{fig:mhdwave_smr_conv} shows the $L_1$ and 
$L_\infty$ norms of the dust density errors as a function of $N_x$. 
The overall errors evaluated by $L_1(\rhod)$ 
show little dependence on the method and exhibit nearly second-order convergence for all cases.
This indicates that
the artificial structures observed in \texttt{HB22} and \texttt{Split-Flux} 
are localized near the refinement boundaries and do not influence the overall errors.

The behavior of $L_\infty(\rhod)$ for \texttt{HB22} and \texttt{HLLgd} does not 
significantly depend on $\tstopd$.
\texttt{HLLgd} exhibits second-order convergence, whereas \texttt{HB22} exhibits 
approximately first-order convergence.
The $L_\infty$ norms obtained with \texttt{HB22} are larger than those obtained with 
\texttt{HLLgd} because 
\texttt{HB22} suffers from artificial structures near the refinement boundaries, whereas 
\texttt{HLLgd} does not, as shown in Figure \ref{fig:mhdwave_smr_cmap}.

By contrast, the \texttt{Split-Flux} method exhibits a clear transition from the \texttt{HLLgd}-type behavior 
to the \texttt{HB22}-type behavior.
In the strong-coupling case ($\tstopd c_\mathrm{ph} k = 10^{-2}$), 
the $L_\infty(\rhod)$ norms obtained with \texttt{Split-Flux} 
behave similarly to those obtained with 
\texttt{HLLgd}.
This is consistent with the fact that the mass fluxes of \texttt{Split-Flux} and \texttt{HLLgd} are 
consistent with the corresponding gas mass fluxes.
The $L_\infty$ norms of \texttt{Split-Flux} are slightly smaller than those of \texttt{HLLgd}.
For $\tstopd c_\mathrm{ph} k = 1$, 
the $L_\infty(\rhod)$ norms obtained with \texttt{Split-Flux} are nearly identical to
those obtained with \texttt{HB22}, as expected.

The artificial structures that emerge near the refinement boundaries in the marginal- and weak-coupling 
regimes are a consequence of the intended limiting behavior of \texttt{Split-Flux}.
When the dust--gas drift is resolved by the grid, 
the dust flux should reduce to 
the pressureless-dust flux, while the gas-carried flux is mainly needed when 
the dust--gas relative motion is unresolved.
Suppressing artifacts near the refinement boundaries in the resolved pressureless-dust regime
may require additional numerical diffusion, as in \texttt{HLLgd}.
This issue is separate from the unresolved strong-coupling problem addressed by
\texttt{Split-Flux} and is beyond the scope of this paper.

\subsection{Kelvin-Helmholtz Instability with Adaptive Mesh Refinement}\label{sec:kh}

To assess the performance of \texttt{Split-Flux} with adaptive mesh refinement (AMR),  
we perform a Kelvin-Helmholtz instability test based on \citet{Stone2020ApJS..249....4S} and 
\citet{HuangBai2022ApJS..262...11H}.

The simulation box is $-0.5\le x,y \le 0.5$. 
The two initial shear layers are centered at $y=\pm 0.25$.
The initial condition is given by 
\begin{equation}
    \rhog(x,y) = \frac{3}{2} - \frac{1}{2}\tanh\left(\frac{ |y| - 0.25}{L}\right),
\end{equation}
\begin{equation}
    v_{\mathrm{g}x}(x,y) = \frac{1}{2}\tanh\left(\frac{ |y| - 0.25}{L}\right),
\end{equation}
\begin{equation}
    v_{\mathrm{g}y}(x,y) = A\sin(4\pi x)\exp\left\{ - \frac{(|y|-0.25)^2}{\sigma^2}\right\},
\end{equation}
where $L=0.01$, $\sigma=0.2$, and $A=0.01$. 
The initial magnetic field is aligned 
to the $x$-axis and its strength is $0.1$.
The gas pressure is spatially constant, $\Pg=2.5$.
We adopt an adiabatic index of $\gamma=1.4$.
The initial dust-to-gas mass ratio is set to $f_\mathrm{dg,ini}=10^{-3}$.
The initial dust velocity is identical to the gas velocity.
The periodic boundary conditions are imposed in all directions. 

We perform both uniform-grid and AMR simulations.
For the uniform-grid runs, $2048\times 2048$ cells are considered.
The AMR runs use a $256\times 256$ base grid 
with \texttt{MeshBlocks} of size $8\times 8$ in the base level (level 0). 
The \texttt{MeshBlocks} 
are refined up to level 3, indicating 
that the effective resolution of the AMR runs is the same as the resolution of the uniform-grid runs.
We define the dimensionless refinement indicator based on 
the cell-scale velocity jump,
\begin{equation}
    g = \Delta x\max\left(
    \Bigr|\frac{\partial v_{\mathrm{g}y}}{\partial x}
    \Bigr|,
    \Bigr|\frac{\partial v_{\mathrm{g}x}}{\partial y}
    \Bigr|,
    \Bigr|\frac{\partial v_{\mathrm{d}y}}{\partial x}
    \Bigr|,
    \Bigr|\frac{\partial v_{x\mathrm{d}}}{\partial y}
    \Bigr|
    \right),
\end{equation}
where $\Delta x$ is the cell width.
A \texttt{MeshBlock} is refined 
if $g$ is larger than $0.01$ 
or derefined if $g$ is smaller than $0.005$.

\begin{figure}[htpb]
    \centering
    \includegraphics[width=1.0\linewidth]{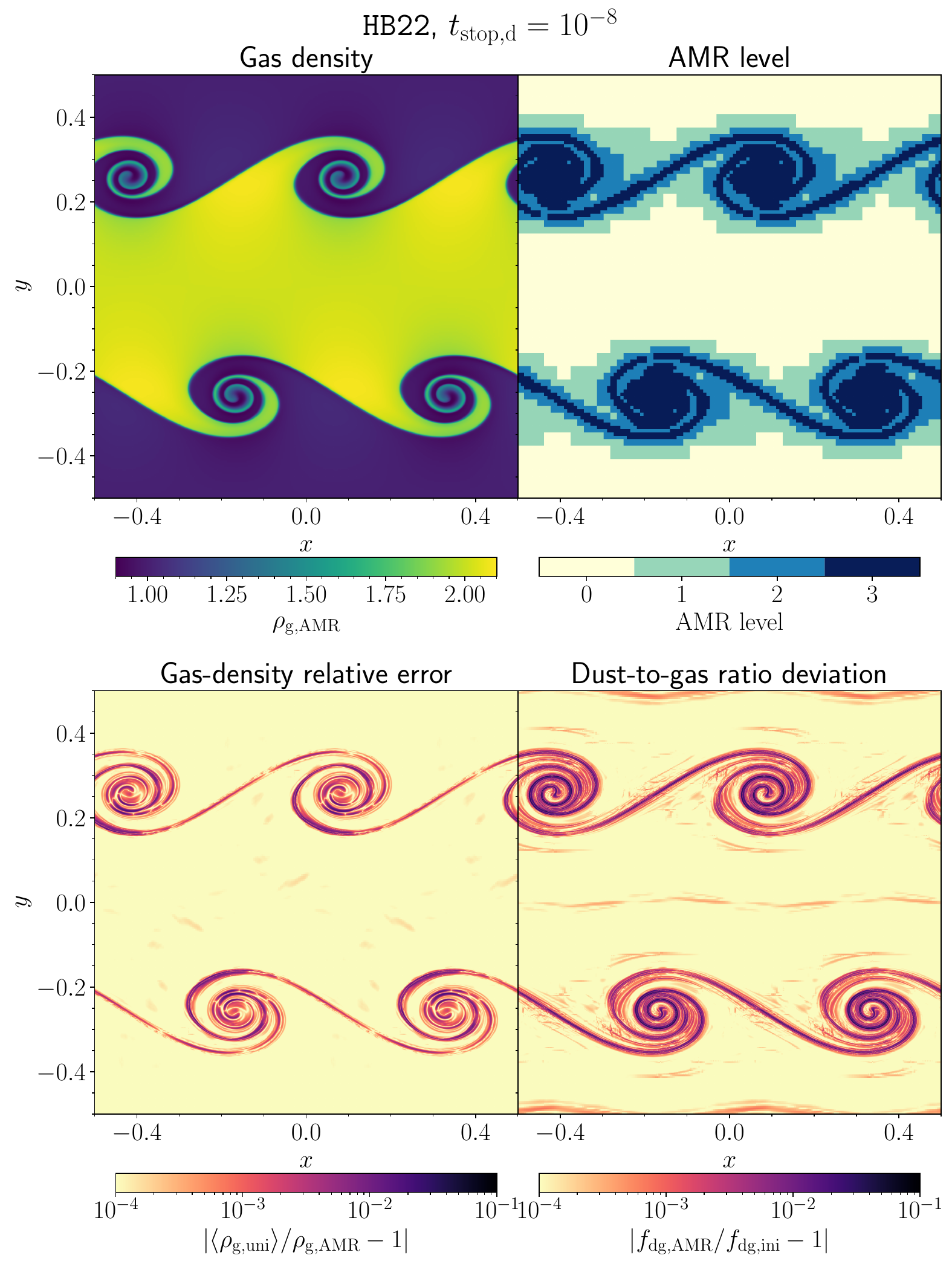}
    \caption{
    Results of the Kelvin-Helmholtz instability at $t=1.5$ for 
    \texttt{HB22} with $t_\mathrm{stop,d}=10^{-8}$.
    The top left and right panels show the gas density in the AMR run ($\rho_\mathrm{g,AMR}$) and 
    the AMR level, respectively.
    The lower left panel shows the relative difference in the gas densities 
    between the AMR and uniform-grid runs, where $\langle \rho_\mathrm{g,uni}\rangle$ stands for the gas density in the 
    uniform-grid run restricted and volume-averaged to the corresponding AMR grids.
    The lower right panel shows the deviation of the dust-to-gas mass ratio 
    in the AMR run from the initial value.
    }
    \label{fig:khhb}
\end{figure}

\begin{figure}[htpb]
    \centering
    \includegraphics[width=1.0\linewidth]{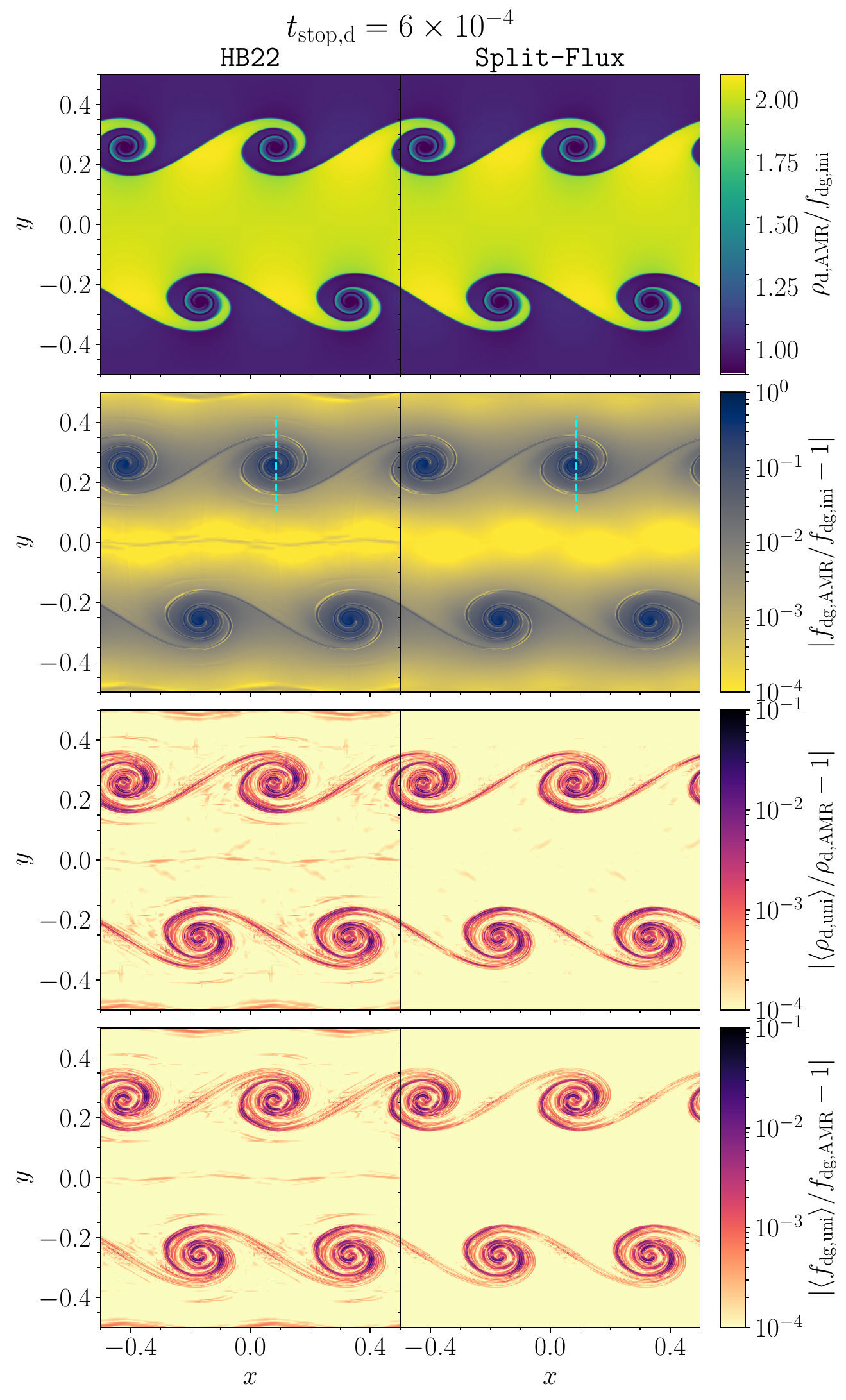}
    \caption{
    Results of the Kelvin-Helmholtz instability at $t=1.5$ 
    for \texttt{HB22} (left) and \texttt{Split-Flux} (right) with 
    AMR. The dust stopping time is $\tstopd = 6\times 10^{-4}$.
    The first and second rows show the scaled dust density 
    $\rhod/f_\mathrm{dg,ini}$ and the deviation of $f_\mathrm{dg}$ from the initial value, respectively. 
    The third and fourth rows represent 
    the fractional differences of ($\rho_\mathrm{d}$, $f_\mathrm{dg}$)
    between the AMR runs and uniform-grid runs, respectively.
    $\langle A_\mathrm{uni}\rangle$ shows a variable $A$ obtained 
    in the uniform-grid runs that is restricted and volume-averaged to the corresponding AMR cells. 
    In the second row, the vertical dashed lines correspond to 
    the segments showing one-dimensional profiles in Figure \ref{fig:kh1d}.
    }
    \label{fig:kh2dmap}
\end{figure}

\begin{figure*}[htpb]
    \centering
    \includegraphics[width=1.0\linewidth]{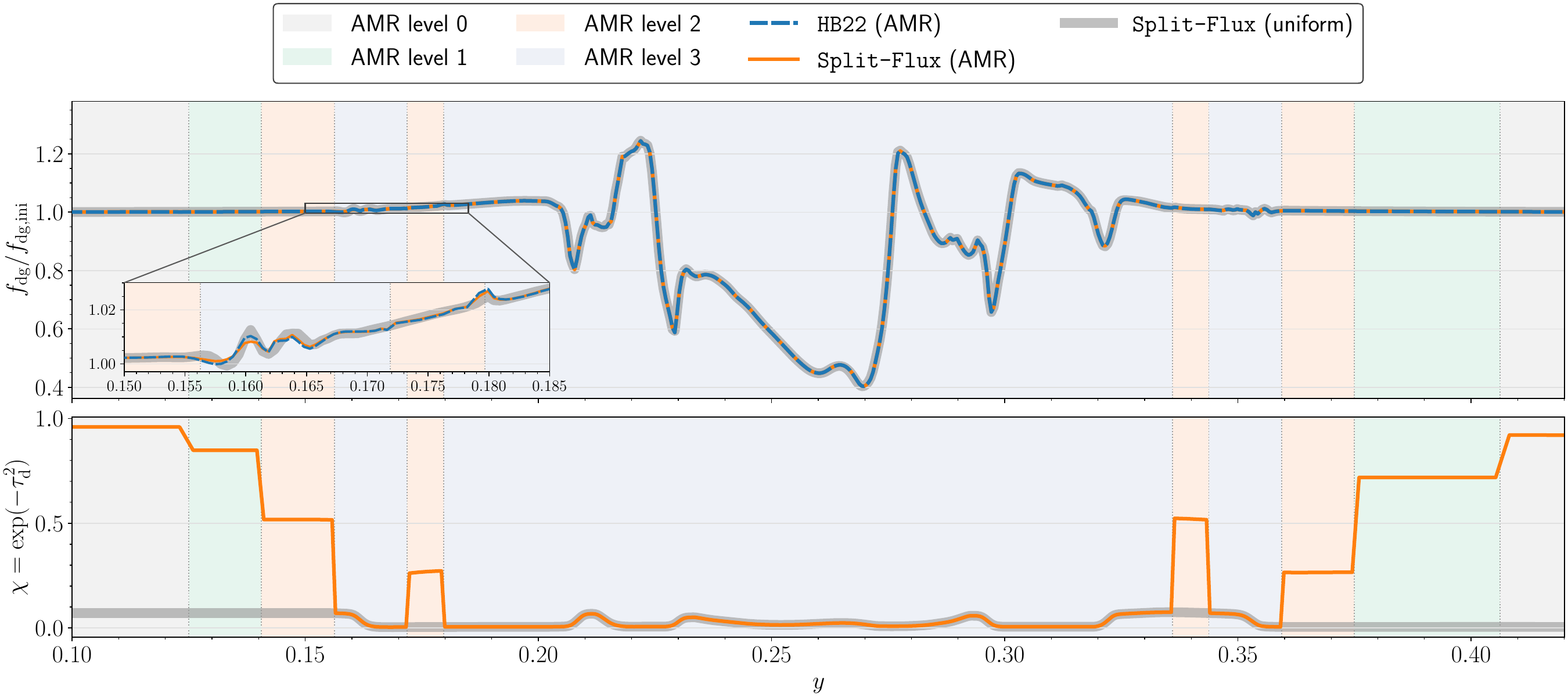}
    \caption{
    Profiles of $f_\mathrm{dg}$ (top)
    and $\chi$ (bottom) at $t=1.5$ along 
    $x=0.087$ for $\tstopd=6\times 10^{-4}$.
    The dashed and solid lines show the results obtained with 
    the AMR \texttt{HB22} and \texttt{Split-Flux} runs, respectively.
    The thick gray lines correspond to the uniform-grid \texttt{Split-Flux} runs.
    The dotted vertical lines  show the refinement boundaries.  
    The $y$ range of each AMR level is shown by different colors.
    }
    \label{fig:kh1d}
\end{figure*}

We consider two different values of 
$\tstopd$.
For the unresolved strong-coupling regime, we consider $\tstopd=10^{-8}$. 
For $\rhog=1$, $\taud$ ranges from $\sim 5\times 10^{-6}$ at level 0 
to $4\times 10^{-5}$ at level 3.
Thus, the dust-gas drift remains unresolved at all refinement levels.
The second value of $\tstopd$ is chosen so that $\chi$ changes 
significantly across refinement levels 0--3.
We set $\tstopd=6\times 10^{-4}$, 
for which $\chi\sim 0.5$ at $\Delta x = 
2^{-9.5}$ and $\rhog=1$.
$\chi$ varies from $0.92$ at level 0 to $5\times 10^{-3}$ at level 3.

We first examine the gas dynamics, which is not affected by 
the dust dynamics because $f_\mathrm{dg}$ is sufficiently small.
The AMR result obtained with \texttt{HB22} for $\tstopd=10^{-8}$ is shown in 
Figure \ref{fig:khhb}.
The finest \texttt{MeshBlocks} 
follow the shear layers and cover the cat's eye vortices.
The lower left panel of Figure \ref{fig:khhb} 
shows the relative differences of the gas density between 
the AMR and uniform-grid runs.
They are concentrated around the density jump, and 
the maximum fractional difference remains $\sim 2\%$.

The fractional deviations of $f_\mathrm{dg}$ 
from the initial value 
for the AMR \texttt{HB22} run with $\tstopd=10^{-8}$ are
concentrated around the density jump
(the lower right panel of Figure \ref{fig:khhb}).
The maximum deviation is around $5\%$. 
For the AMR \texttt{Split-Flux} run, the deviation remains at a
low level of $\sim 10^{-5}$.

For $\tstopd=6\times 10^{-4}$,
reduced dust densities are visible near the cat's-eye cores, together 
with significant departures of $f_\mathrm{dg}$ from the initial value
in the first and second rows 
of Figure \ref{fig:kh2dmap}.
The large-scale distributions of both $\rhod$ and $f_\mathrm{dg}$ are similar 
between \texttt{HB22} and \texttt{Split-Flux}, whereas 
the artificial small-scale structures visible in \texttt{HB22} are 
reduced in \texttt{Split-Flux}.

A more detailed comparison 
of $f_\mathrm{dg}$ between the \texttt{HB22} and \texttt{Split-Flux} 
runs with AMR is shown in Figure \ref{fig:kh1d}.
The upper panel shows the $f_\mathrm{dg}$ profiles along $x=0.087$,
which is the $x$ coordinate of a local gas-pressure minimum 
at $t=1.5$ in the upper-right quadrant.
The results for the \texttt{HB22} and \texttt{Split-Flux} 
runs with AMR are broadly consistent.
The overall \texttt{Split-Flux} profile remains broadly consistent with the uniform-grid solution 
despite abrupt changes in $\chi$ across refinement boundaries.
Small oscillatory features remain near some refinement boundaries, particularly where $\chi$ becomes very small on the finer level 
(see $y\sim 0.18$ in the enlarged view of Figure \ref{fig:kh1d}). 
In such regions, the \texttt{Split-Flux} method locally approaches the \texttt{HB22} pressureless flux, 
whose lower numerical diffusion allows small $\rhod$ perturbations generated near refinement boundaries to persist.
Nevertheless, the amplitude of these features remains small.
Around $y\simeq0.16$, the \texttt{Split-Flux} solution with AMR slightly underestimates 
the amplitudes of small-scale variations relative to the uniform-grid solution and the \texttt{HB22} result with AMR, 
indicating additional numerical diffusion from the coupled-limit contribution. This damping is modest and does not produce spurious amplification.

The third and fourth rows of Figure \ref{fig:kh2dmap} show 
the fractional differences in 
$\rhod$ and $f_\mathrm{dg}$, respectively, between the uniform-grid and
AMR runs for \texttt{HB22} and \texttt{Split-Flux}.
The peak differences are comparable between the two methods, 
reaching $\sim 4\%$ for $\rhod$ and $\sim 3\%$ for $f_\mathrm{dg}$.
These differences are concentrated around the nonlinear vortices.
The \texttt{Split-Flux} method yields smaller differences than \texttt{HB22} in the other regions.

\begin{figure}[htpb]
    \centering
    \includegraphics[width=1.0\linewidth]{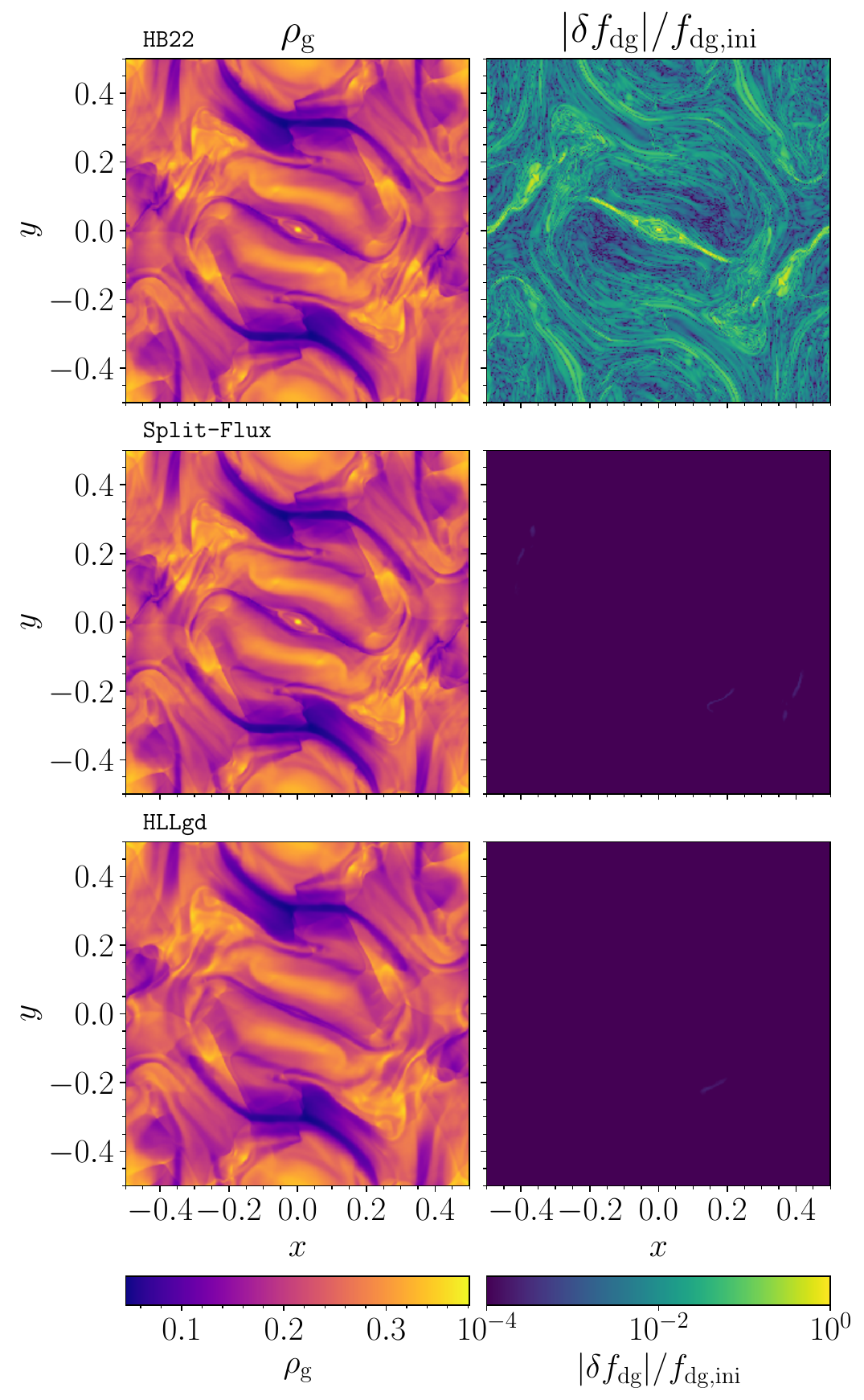}
    \caption{
    Results of the Orszag--Tang vortex test at $t=1$ 
    for \texttt{HB22} (top), 
    \texttt{Split-Flux} (middle), and \texttt{HLLgd} (bottom).
    The results with a resolution of $512\times 512$ and $\tstopd=10^{-9}$ are shown.
    The left panels show the gas density, and 
    the right panels show the absolute fractional deviation of $f_\mathrm{dg}$ from its initial value,
    $|\delta f_\mathrm{dg}|/f_\mathrm{dg,ini}=
    |f_\mathrm{dg}/f_\mathrm{dg,ini}-1|$.
    }
    \label{fig:ot_map}
\end{figure}

\subsection{Orszag--Tang Vortex}\label{sec:ot}

The Orszag--Tang vortex is a standard MHD test problem in which
two-dimensional MHD turbulence develops from a simple initial condition
\citep{OrszagTang1979JFM....90..129O}.
Because it contains various MHD waves and discontinuities,
this test is useful for assessing the performance of dust solvers
in turbulent flows.

The initial state consists of a uniform gas with
$\rho_\mathrm{g0}=25/(36\pi)$ and $P_\mathrm{g0}=5/(12\pi)$.
The initial velocity and magnetic field are given by
$(v_{x\mathrm{g}0},v_{y\mathrm{g}0})=(-\sin (2\pi y),\sin (2\pi x))$ and
$\bm{B}_0 = (B_{x0},B_{y0})=(-\sin (2\pi y),\sin (4\pi x))/\sqrt{4\pi}$, respectively.
We adopt an adiabatic index of $\gamma=5/3$.
The simulation domain is $-0.5\le x,y \le 0.5$, and
periodic boundary conditions are imposed in both directions.
The initial dust density is set to $10^{-3}\rho_\mathrm{g0}$, corresponding to
$f_\mathrm{dg,ini}=10^{-3}$.
The initial dust velocity is identical to the gas velocity.
Since $f_\mathrm{dg,ini}$ is extremely small, 
the gas dynamics is not affected by the dust in this test.
The adopted resolution is $512 \times 512$.

We first consider the cases with $\tstopd=10^{-9}$.
The initial reference value of $\taud$ is 
$\tau_\mathrm{d0} = c_\mathrm{f0}\tstopd/\Delta x \sim 6\times 10^{-7}$, 
where $c_\mathrm{f0} = \sqrt{(\gamma P_\mathrm{g0} + \langle \bm{B}_0^2\rangle)/\rho_\mathrm{g0}}$, and 
$\langle \bm{B}_0^2\rangle$ denotes the volume average of $\bm{B}_0^2$,
indicating that 
the gas and dust are expected to be tightly coupled during the evolution.

Figure \ref{fig:ot_map} compares the gas density and the relative 
deviation of the dust--to--gas mass ratio from the initial value.
The differences among the gas density maps arise mainly from the differences
in the gas solvers\footnote{ 
In the central region, the plasmoids form in the \texttt{HB22} and 
\texttt{Split-Flux} runs, whereas no comparable structures appear in the \texttt{HLLgd} run.
We do not focus on the detailed reconnection structures around the 
center in the Orszag--Tang vortex, 
which can depend on resolution and numerical
dissipation. Instead, we use this problem as a turbulent MHD test to compare the dust transport among the different solvers.
}.

The effect of the dust solver is clearly seen in the relative
deviation of the dust--to--gas mass ratio from the initial value,
$|\delta f_\mathrm{dg}|/f_\mathrm{dg,ini}$, where $\delta f_\mathrm{dg} = f_\mathrm{dg} - f_\mathrm{dg,ini}$.
Although the dust and gas are expected to be tightly coupled 
for such a small stopping time $\tstopd$, the \texttt{HB22} solver produces spatial variations in the dust--to--gas mass ratio.
The top-right panel of Figure~\ref{fig:ot_map} shows that
$|\delta f_\mathrm{dg}|/f_\mathrm{dg,ini}$ reaches values of order unity.
By contrast, $|\delta f_\mathrm{dg}|/f_\mathrm{dg,ini}$ remains 
small in \texttt{Split-Flux} and \texttt{HLLgd}.
With \texttt{Split-Flux}, the $L_2$ norm of
$|\delta f_\mathrm{dg}|/f_\mathrm{dg,ini}$ remains at $\sim 10^{-5}$,
close to the value of $\sim 8\times 10^{-6}$ obtained with
\texttt{HLLgd}.

\begin{figure}[htpb]
    \centering
    \includegraphics[width=0.8\linewidth]{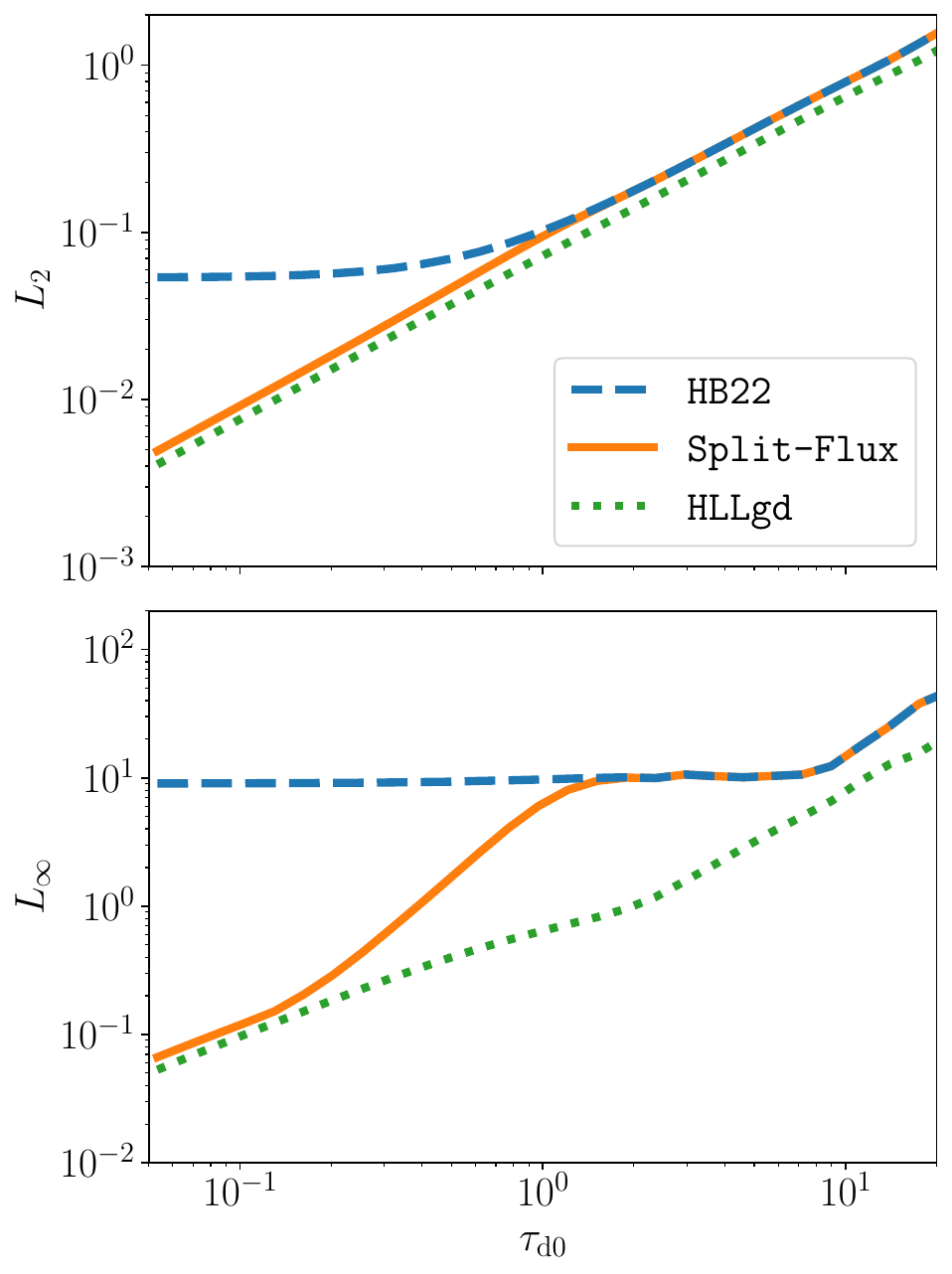}
    \caption{
    $L_2$ (top) and $L_\infty$ (bottom)
    norms of $|f_\mathrm{dg}/f_\mathrm{dg,ini}-1|$ averaged over $1\le t \le 1.5$ 
    in the Orszag--Tang vortex 
    test for \texttt{HB22} (dashed), 
    \texttt{Split-Flux} (solid), and \texttt{HLLgd} (dotted).
    The horizontal axes show
    $\tau_\mathrm{d0} = c_\mathrm{f0}t_\mathrm{stop,d}/\Delta x$,
    where $c_\mathrm{f0} = \sqrt{(\gamma P_\mathrm{g0} + \langle \bm{B}_0^2\rangle)/\rho_\mathrm{g0}}$, and 
    $\langle \bm{B}_0^2\rangle$ denotes the volume average of $\bm{B}_0^2$.
    }
    \label{fig:ot_K_dfdg}
\end{figure}

\begin{figure*}[htpb]
    \centering
    \includegraphics[width=1.0\linewidth]{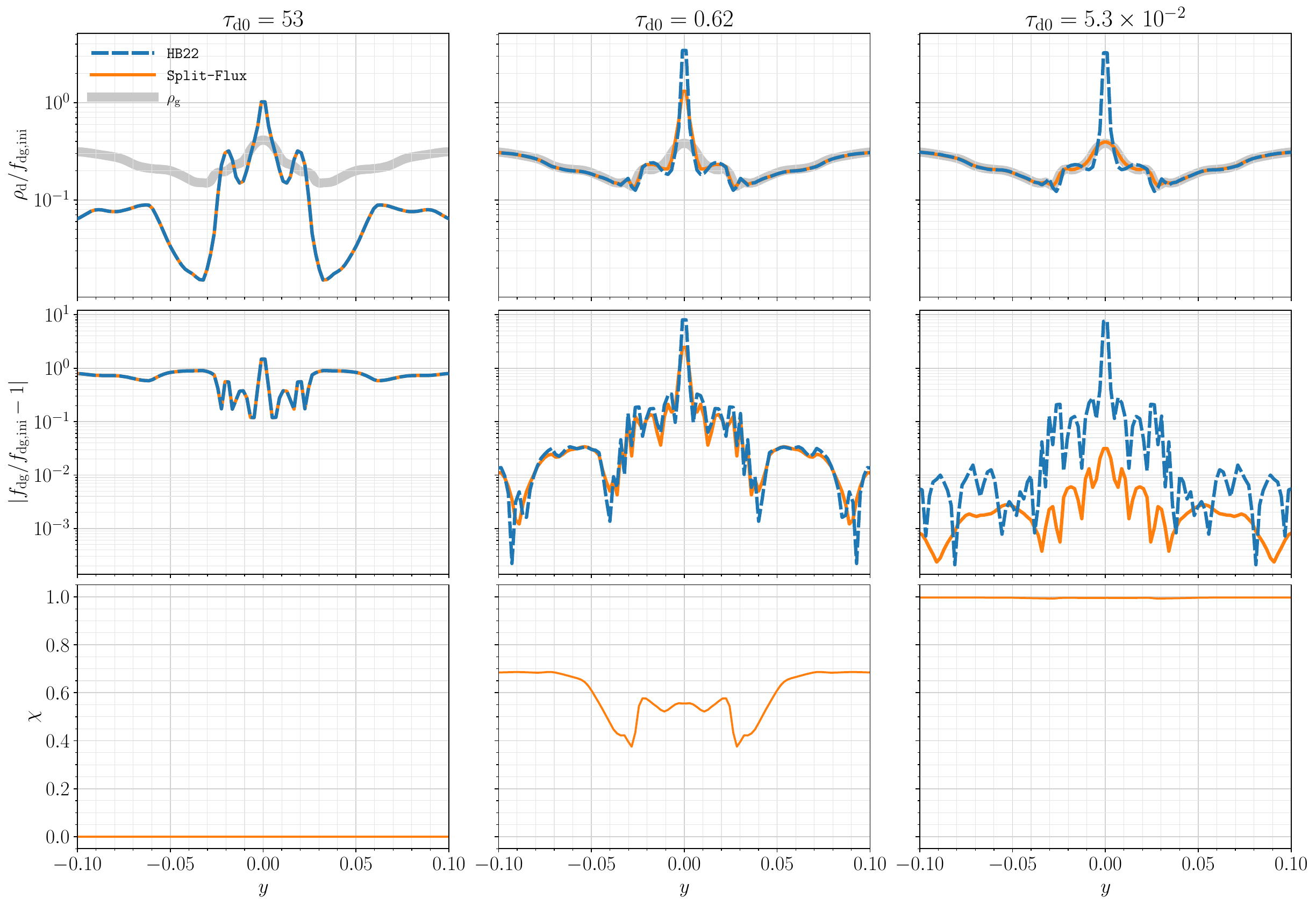}
    \caption{
    Profiles of $\rhod/f_\mathrm{dg,ini}$ (top), $|f_\mathrm{dg}/f_\mathrm{dg,ini}-1|$ (middle) and 
    $\chi$ (bottom) at $t=1$ along $x=0$ for three different values of 
    $\tau_\mathrm{d0}$: $53$, $0.62$, and $5.3\times 10^{-2}$.
    The dashed and solid lines show the results obtained with
    \texttt{HB22} and \texttt{Split-Flux}, respectively.
    In the top panels, the gas densities are shown by the gray 
    lines.
    The transition function $\chi$ shown in the bottom panels 
    is evaluated at cell centers by taking the larger of the 
    arithmetic averages of the face-centered values in 
    the $x$ and $y$-directions,
    $\max[\{ (\chi)_{i-\frac{1}{2},j,k}
           + (\chi)_{i+\frac{1}{2},j,k}
    \}/2,
    \{ (\chi)_{i,j-\frac{1}{2},k}
           + (\chi)_{i,j+\frac{1}{2},k}
    \}/2]$.
    }
    \label{fig:ot_1d}
\end{figure*}

To examine the dependence of $|\delta f_\mathrm{dg}|/f_\mathrm{dg,ini}$ on $\tstopd$, 
we conduct simulations of 
the Orszag--Tang vortex with various values of $\tstopd$ and 
compute the $L_2$ and $L_\mathrm{\infty}$ norms of 
$|\delta f_\mathrm{dg}|/f_\mathrm{dg,ini}$ 
averaged over $1\le t\le 1.5$.
Figure \ref{fig:ot_K_dfdg} compares the time-averaged $L_2$ and $L_\mathrm{\infty}$ norms 
for \texttt{HB22}, \texttt{Split-Flux}, and \texttt{HLLgd}.
The horizontal axes show the dimensionless dust stopping time
$\tau_\mathrm{d0}$ evaluated from the initial conditions.

We first examine the $L_2$ norms, which measure the overall deviation
of the dust--to--gas mass ratio from the initial value.
All solvers show similar behavior when $\tau_\mathrm{d0} \gtrsim 1$, and  
the $L_2$ norms increase as $\tstopd$ increases in this regime because 
the coupling becomes progressively weaker.
By contrast, when $\tau_\mathrm{d0}$ decreases below unity, 
the $L_2$ norms for \texttt{HB22} no longer decrease even
though the dust--gas coupling becomes stronger.
The \texttt{Split-Flux} and \texttt{HLLgd} solvers 
continue to reduce the $L_2$ norms with decreasing $\tau_\mathrm{d0}$.
This demonstrates that both solvers substantially 
suppress the spurious 
dust--to--gas mass ratio variations that appear in \texttt{HB22}
in the strong-coupling regime.

We next examine the $L_{\infty}$ norms, which are sensitive to 
localized variations of $|\delta f_\mathrm{dg}|/f_\mathrm{dg,ini}$.
For \texttt{HB22}, the $L_{\infty}$ norms do not approach zero
with decreasing $\tau_\mathrm{d0}$, and exceed unity even in the 
strong-coupling regime, $\tau_\mathrm{d0}\ll 1$, 
as seen in the top-right panel of Figure \ref{fig:ot_map}.
The $L_\infty$ norm obtained with 
\texttt{Split-Flux} deviates from 
that obtained with \texttt{HB22} for $\tau_\mathrm{d0}\lesssim 1$ and 
continues to decrease with decreasing $\tau_\mathrm{d0}$. 
This shows that \texttt{Split-Flux} also reduces
large local deviations in the strong-coupling regime.

Among the three solvers, \texttt{HLLgd} gives the smallest
$L_\infty$ values over the range shown here.
Unlike the $L_2$ norms, the differences between the $L_\infty$ norms obtained with \texttt{HLLgd} 
and \texttt{Split-Flux} are relatively large for 
$\tau_\mathrm{d0}\gtrsim 0.1$ because 
the $L_\infty$ norm depends more strongly on localized structures.
The difference between \texttt{Split-Flux} and \texttt{HLLgd} may 
reflect both the difference between the gas solvers, 
HLLD and HLL,
and the different levels of numerical dissipation introduced in the dust solvers for $\tau_\mathrm{d0}\gtrsim 1$.

Figure \ref{fig:ot_1d} compares one-dimensional profiles along $x=0$.
We select this cut because, in the \texttt{HB22} run with $\tstopd=10^{-9}$, the maximum value of 
$|\delta f_\mathrm{dg}|/f_\mathrm{dg,ini}$ occurs along this line.
For $\tau_\mathrm{d0}=53$, the $\rhod$ profile obtained with 
\texttt{Split-Flux} is nearly identical to that obtained with 
\texttt{HB22} because $\chi$ is almost zero.

Next, we examine the case with $\tau_\mathrm{d0} =0.62$, near which
the $L_2$ norms obtained with \texttt{HB22} and \texttt{Split-Flux} 
begin to diverge in Figure \ref{fig:ot_K_dfdg}.
The $\rhod$ profiles obtained with \texttt{Split-Flux} and \texttt{HB22} remain broadly similar, 
although small deviations appear around the center, where
a current sheet exists.
A noticeable mismatch between 
$\rhod/f_\mathrm{dg,ini}$ and $\rhog$ appears only near the central current sheet at 
$y\sim 0$.
For $|y|\gtrsim 0.01$, $\chi$ takes intermediate values ($\chi\sim 0.4-0.7$), indicating 
that both the coupled-limit and pressureless fluxes contribute to the dust flux.
No artificial structures caused by variations of $\chi$ are seen.

In the strong-coupling case ($\tau_\mathrm{d0}=5.3\times 10^{-2}$), $\chi$ is almost unity, 
indicating that the dust flux is dominated by $(\bm{F}_\mathrm{d})_\mathrm{cpl}$
(Equation (\ref{splitflux})).
For \texttt{HB22}, significant errors are seen in $|y|\lesssim 
0.04$. These errors are substantially reduced by \texttt{Split-Flux}, and 
$\rhod/f_\mathrm{dg,ini}$ agrees well with $\rhog$ over 
the entire profile.
For $|y|>0.05$, \texttt{HB22} also produces only small deviations
in $f_\mathrm{dg}$.
This is consistent with the behavior discussed in Section~\ref{sec:shock}.
The errors in $f_\mathrm{dg}$ are concentrated around the sharp structures associated with 
the central current sheet and nearby shocks, whereas they remain small in the outer smoothly 
varying regions.

\section{Discussion} \label{sec:discussion}

\subsection{Streaming Instability}

Streaming instability (SI) is a promising mechanism for
triggering planetesimal formation
\citep{YoudinGoodman2005ApJ...620..459Y}.
It is useful to examine whether 
numerically resolved SI modes can lie in the regime $\taud \lesssim 1$, where 
the \texttt{Split-Flux} method differs from the conventional pressureless dust solver.

As shown in Appendix \ref{app:SI}, the SI modes satisfying $\taud\lesssim 1$ formally exist.
However, in the gas-dominated regime, such modes have extremely small growth rates. 
Therefore, we expect that SI modes that grow to dynamically significant amplitudes in practical simulations 
generally have $\taud\gtrsim 1$. In this regime, $\chi = \exp(-\taud^2)$ is small, and 
the \texttt{Split-Flux} dust flux approaches the pressureless dust flux.
In the dust-dominated regime \citep{YoudinGoodman2005ApJ...620..459Y,SquireHopkins2018MNRAS.477.5011S}, 
the high-$f_\mathrm{dg}$ modes with $\taud \lesssim 1$ can exist when 
the radial wavenumber is much larger than the vertical wavenumber.
Such highly anisotropic modes have small growth rates proportional to $K_z/K_x$, 
and their high radial wavenumbers may make them sensitive to physical diffusion 
and viscosity, where $K_x$ and $K_z$ are the dimensionless radial and vertical wavenumbers defined in Appendix \ref{app:SI}.

In summary, the \texttt{Split-Flux} method is not expected to significantly modify the linear 
development of these SI modes.

\subsection{Extensions to Multifluid Dust Description}

The \texttt{Split-Flux} method can be straightforwardly extended to a multifluid description of dust.
In such approaches \citep{Benitez-Llambay2019ApJS..241...25B,HuangBai2022ApJS..262...11H},
the continuity and momentum equations of each species have independent 
advection fluxes, while the dust species are coupled through their interaction with the gas.
The \texttt{Split-Flux} fluxes of the $s$-th dust species are given by 
\begin{equation}
    \bm{F}_{\mathrm{d},s} =
    \chi(\tau_{\mathrm{d},s}) 
    (\bm{F}_{\mathrm{d},s})_{\mathrm{cpl}} 
    + \left\{ 1 - \chi(\tau_{\mathrm{d},s}) \right\} 
    (\bm{F}_{\mathrm{d},s})_{\text{\texttt{HB22}}},
    \label{splitflux_multi}
\end{equation}
where 
\begin{equation}
   \tau_{\mathrm{d},s} = \max\left(|\bm{v}_{\mathrm{d},s} - \vvg|,c_{\mathrm{f}n}\right)\frac{t_{\mathrm{stop,d},s}}{\Delta x}.
\end{equation}
Here, $c_{\mathrm{f}n}$ is the fast-wave phase speed normal to the cell interface.
The coupled-limit flux, $(\bm{F}_{\mathrm{d},s})_\mathrm{cpl}$, and the \texttt{HB22} flux, 
$(\bm{F}_{\mathrm{d},s})_{\text{\texttt{HB22}}}$, of the $s$-th dust species are computed using 
Equations (\ref{hb22}), (\ref{Fdg}), and (\ref{Fdrel}) by replacing all dust quantities 
with those of species $s$.

In Appendix \ref{app:kh_multi}, we present a KHI test with two dust species following 
\citet{HuangBai2022ApJS..262...11H}.
One is a strongly coupled dust species for which the dust--gas drift is unresolved, and 
the other is a marginally coupled dust species.
We confirmed that the \texttt{Split-Flux} method keeps deviations of $f_{\mathrm{dg},s} = \rho_{\mathrm{d},s}/\rhog$ from its initial value sufficiently small 
for the strongly coupled dust species, while reproducing the \texttt{HB22} result for the marginally coupled species.
This demonstrates that the \texttt{Split-Flux} method can simultaneously recover the appropriate 
coupled- or pressureless-flux limit for each dust species according to its stopping time.

\section{Summary} \label{sec:summary}

In dust--gas multifluid systems, 
previous studies discussed the importance of resolving the 
stopping length, which is estimated as $c_\mathrm{s}\tstopd$ without 
magnetic fields, to capture dust--gas drift accurately.
If the stopping length is not resolved,
artificial disturbances in the dust--to--gas mass ratio 
$f_\mathrm{dg}$ can be generated when 
independent numerical fluxes are used to update the dust and gas variables.
\citet{Verrier2025A&A...701A.174V} developed a new method, \texttt{HLLgd}, in which 
consistency between the dust and gas mass fluxes is enforced by constructing both fluxes within 
the HLL framework.
The numerical fluxes of \texttt{HLLgd} are applicable across regimes from strong to weak coupling.
\citet{Verrier2025A&A...701A.174V} showed that \texttt{HLLgd} significantly reduces artificial variations 
in $f_\mathrm{dg}$ in the strong-coupling regime.

Straightforward extensions of the \texttt{HLLgd}-type approach to less dissipative Riemann solvers, such as HLLD for MHD,  
are difficult because the HLLD fluxes are constructed from intermediate states determined by the 
gas dynamics, including gas pressure and Maxwell stress.
These intermediate-state constructions cannot be directly transferred 
to a pressureless dust solver 
because they depend on gas pressure and Maxwell stress.
The applicability of \texttt{HLLgd} relies on an important property of the HLL method.
The HLL flux can be expressed using only the left and right states and the two signal speeds, without 
resolving the detailed intermediate-wave structure.

Our strategy differs from that of \texttt{HLLgd} (Section \ref{sec:newmethod}).
In the strong-coupling regime, both dust mass and momentum 
should be transported by the gas-carried flux associated with the gas mass flux.
In the weak-coupling regime, however, the dust
should recover pressureless dynamics,
because using a dust flux consistent with the gas mass flux 
would artificially introduce the effects of gas dynamics into the dust dynamics even 
when the dust and gas are decoupled.
To recover these two limiting behaviors, 
the newly developed \texttt{Split-Flux} method switches 
smoothly between a dust flux consistent with the HLLD gas mass 
flux and the pressureless \texttt{HB22} dust flux \citep{HuangBai2022ApJS..262...11H}, 
using a transition function based on
whether the dust--gas drift is resolved on the grid
(Equation (\ref{splitflux})).

The \texttt{Split-Flux} method uses the dimensionless dust stopping-time parameter
$\taud = S_\mathrm{char}\tstopd/\Delta x$, which compares the characteristic stopping length with the grid size.
The characteristic velocity $S_\mathrm{char}$ is the maximum of 
the dust--to--gas relative speed and the fast-wave speed. 
Because the fast-wave speed is the largest local MHD characteristic speed relative to the gas, it may exceed the 
velocity scale associated with the problems considered, especially for sub-Alfv\'enic motions.

Nevertheless, the numerical experiments also suggest that \texttt{Split-Flux} is not highly 
sensitive to a moderate overestimate of $\tau_\mathrm{d}$.
Even when the pressureless flux is used down to smaller dust stopping times $\tstopd$, the resultant 
errors in the dust--to--gas mass ratio remain small in smoothly varying regions.
The Sod test with a strong parallel magnetic field 
(Section \ref{sec:sod}) shows that 
the fast-wave speed provides a practical and conservative 
estimate of the transition even in a strongly magnetized 
sub-Alfv\'enic flow. 

We conducted various numerical experiments in Section \ref{sec:experiments} 
and confirmed that the \texttt{Split-Flux} method reduces artificial variations in $f_\mathrm{dg}$ to levels comparable to those obtained with \texttt{HLLgd} in the unresolved strong-coupling regime. 
The \texttt{Split-Flux} method captures gas structures more sharply than \texttt{HLLgd} because
it uses the less dissipative HLLD solver.  

We investigated the behavior of \texttt{Split-Flux} with mesh refinement 
using the wave-propagation test with SMR (Section \ref{sec:smr}) and 
the KHI test with AMR (Section \ref{sec:kh}).
For the SMR test, 
the \texttt{Split-Flux} method does not generate appreciable artificial 
disturbances near the refinement boundaries when dust--gas drift is unresolved 
on both levels.
The KHI test with AMR shows no large boundary artifacts even though 
the transition function $\chi$ varies substantially across the refinement 
levels.
These results demonstrate that the \texttt{Split-Flux} method yields robust results 
with mesh refinement when the coupled-limit flux contributes significantly.

Nevertheless, the \texttt{Split-Flux} method inherits the numerical issues of the 
pressureless dust solver 
when the dust--gas drift is resolved.
For instance, in the wave propagation tests with SMR 
presented in Section \ref{sec:smr}, 
we found that artificial disturbances emerge at the grid scale near the refinement
boundaries in \texttt{Split-Flux} in the marginal-coupling case.
Also in the KHI test (Section \ref{sec:kh}), small artificial wiggles remain near the refinement boundaries where 
$\chi$ becomes very small on the finer level.
These structures arise from the pressureless nature of the dust and are not specific to the gas--consistency problem addressed in 
this paper. Numerical dissipation introduced in \texttt{HLLgd} can reduce such errors.

These results demonstrate that \texttt{Split-Flux} provides a practical way to reduce artificial variations in the dust--to--gas mass 
ratio in the strong-coupling regime while retaining the lower 
numerical diffusivity of the HLLD gas solver.
Although we have focused on HLLD in this work, 
the formulation of \texttt{Split-Flux} is not specific to HLLD and can 
in principle be combined with other gas solvers by constructing the gas-carried component from the corresponding gas mass flux.

\begin{acknowledgments}
This work is supported by Japan Society for the Promotion of Science (JSPS) KAKENHI Grant Number JP22KK0043.
Numerical computations were in part carried out on HPE Cray XD2000 at the Center for Computational Astrophysics, National Astronomical Observatory of Japan.
ChatGPT (model GPT-5.6) was used to assist with checking the internal consistency of the manuscript and with English-language editing. OpenAI Codex (model gpt-5.6-sol) was used to assist with code development. All suggestions and outputs generated by these tools were reviewed and verified by the author, who takes full responsibility for the content of the manuscript and the correctness of the code.
\end{acknowledgments}

{\software{ Athena++ \citep{Stone2020ApJS..249....4S,HuangBai2022ApJS..262...11H}, NumPy \citep{Numpy}, Matplotlib \citep{Matplotlib}, ChatGPT \citep{openai_gpt56_2026},  Codex \citep{openai_codex_2026}}}


\appendix

\section{Different Choices of Transition Function $\chi$}\label{app:chi}

The functional form of the transition function $\chi(\taud)$ is not unique.
Here, we consider the following functional forms:
\begin{equation}
    \chi_1 = \exp(-\tau_\mathrm{d}^2),\;\;\;
    \chi_2 = \frac{1}{1+ \tau_\mathrm{d}^2},
\end{equation}
and 
\begin{equation}
    \chi_3 = \exp(-\tau_\mathrm{d}),\;\;\;
    \chi_4 = \frac{1}{1+ \tau_\mathrm{d}}.
\end{equation}
The first candidate $\chi_1$ is adopted in this paper (Equation (\ref{transitionfunc})).

In the small-$\taud$ limit, 
$\chi_1$ and $\chi_2$ have the same leading-order behavior, 
as do $\chi_3$  and $\chi_4$:
\begin{equation}
\chi_{1,2} = 1-\tau_\mathrm{d}^2 + O(\taud^4)\quad
\mathrm{for}~\taud\rightarrow 0
\end{equation}
and 
\begin{equation}
\chi_{3,4} = 1-\tau_\mathrm{d} + O(\taud^2) \quad
\mathrm{for}~\taud\rightarrow 0
\end{equation}
Thus, for $\taud\ll 1$, the contribution of the pressureless flux $(\bm{F}_\mathrm{d})_{\text{\texttt{HB22}}}$, which 
is proportional to $(1-\chi)$, is 
 larger for $\chi_3$ and $\chi_4$ than for $\chi_1$ and $\chi_2$.
 
In the large-$\taud$ limit, 
$\chi_2$ and $\chi_4$ decay algebraically, whereas $\chi_1$ and $\chi_3$ decay 
exponentially. The coupled-limit flux remains important up to larger values of $\taud$ 
for $\chi_2$ and $\chi_4$.

\begin{figure}[htpb]
    \centering
    \includegraphics[width=1.0\linewidth]{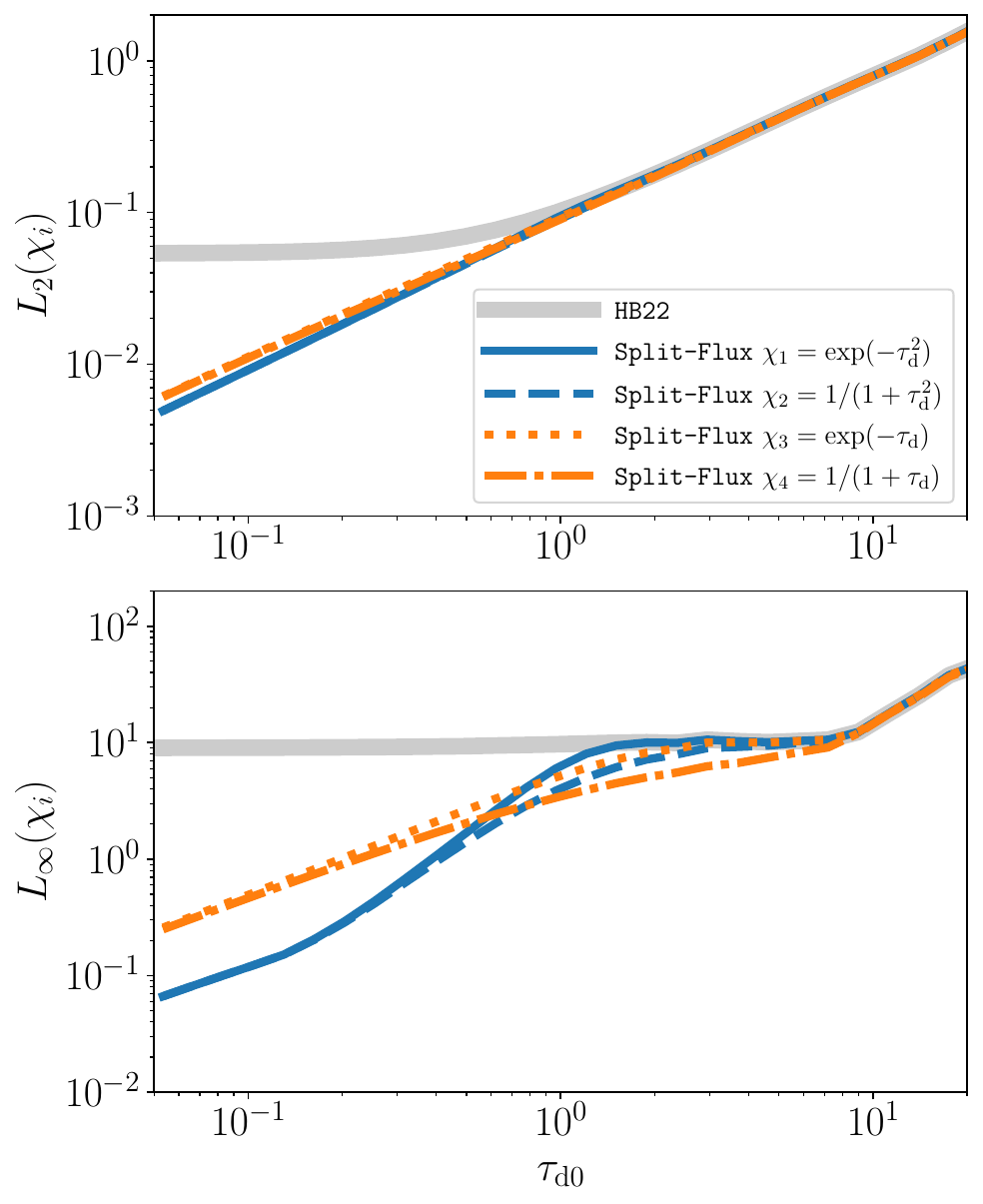}
    \caption{
    $L_2$ (top) and $L_\infty$ (bottom) norms of $f_\mathrm{dg}/f_\mathrm{dg,ini}-1$ averaged over 
    $1\le t\le 1.5$ in the Orszag-Tang vortex test for 
    four choices of the transition function $\chi$.
    The horizontal axes show the initial reference parameter $\tau_\mathrm{d0}$.
    For comparison, the results for \texttt{HB22} are plotted 
    by the thick gray lines.
    }
    \label{fig:ot_K_dfdg_chichoice}
\end{figure}

To compare the performance of the four choices of $\chi$, 
we show the Orszag--Tang results 
in Figure \ref{fig:ot_K_dfdg_chichoice}.
The overall differences among the $L_2$ norms are small.
For $\tau_\mathrm{d0}<1$, 
the $L_2$ norms obtained with $\chi_3$ and $\chi_4$ are
slightly larger than those obtained with $\chi_1$ and $\chi_2$.
This dependence is consistent with their asymptotic behaviors in the small-$\taud$ limit, where 
the contribution of the pressureless 
flux is larger for 
($\chi_3$, $\chi_4$)
than for ($\chi_1$, $\chi_2$).

The $L_\infty$ norm, which is sensitive 
to localized deviations, 
depends more strongly on the functional form of $\chi$,
as seen in the bottom panel of Figure \ref{fig:ot_K_dfdg_chichoice}.
As in the $L_2$ norms, the $L_\infty$ norms for $\chi_3$ and $\chi_4$
are larger than those for $\chi_1$ and $\chi_2$.
Thus, $\chi_1$ and $\chi_2$ yield smaller errors for $\tau_\mathrm{d0} \ll 1$. 

For $1\lesssim \tau_\mathrm{d0}\lesssim 10$, the
$\chi_1$ results are closer to the \texttt{HB22} results than the $\chi_2$ results,
and the same is true of $\chi_3$ relative to $\chi_4$.
This behavior is attributed to the 
exponential decay of 
$\chi_1$ and $\chi_3$.  

From these numerical experiments, we find that 
the specific functional form of 
$\chi$ has a minor effect on the overall $L_2$ norm, although 
the localized deviations estimated 
by the $L_\infty$ norm are moderately sensitive 
to its functional form.
In this paper, we adopt $\chi_1$ because it provides
a relatively sharp transition between $(\bm{F}_\mathrm{d})_\mathrm{cpl}$
and $(\bm{F}_\mathrm{d})_\mathtt{HB22}$ while 
yielding results comparable to those obtained with the other choices.


\section{Streaming Instability}\label{app:SI}

We estimate whether resolved SI modes can satisfy $\tau_d\lesssim1$. 
We use the standard dimensionless parameters ${\rm St}=\Omega t_\mathrm{stop,d}$, 
$K_{x,z}=\eta r k_{x,z}$, and $\Pi=\eta r\Omega/c_s$, where 
$(x,z)$ are the local shearing-box coordinates for radial and vertical directions,
$r$ is the radius of the corotating frame, $\Omega(r)$ is the Keplerian angular velocity, $\eta = -0.5 (d\ln \Pg/d\ln r)(c_\mathrm{s}/(r\Omega))^2$ represents the strength 
of the radial gas pressure gradient compared with the gravitational force from the central star.

For the SI modes developing inside disks, 
the $z$ wavelength should be less than the scale height, or $K_z\gtrsim 2\pi \Pi$, where 
$\Pi = \eta r \Omega/c_\mathrm{s}$.
Assuming that the $z$ wavelength is resolved by $N_z$ cells, one 
obtains
\begin{equation}
    \tau_{z\mathrm{d}} = \frac{\mathrm{St}N_zK_z}{2\pi \Pi}.
\end{equation}
Since $K_z \gtrsim 2\pi \Pi$, $\tau_{z\mathrm{d}}$ should satisfy 
$\tau_{z\mathrm{d}} \gtrsim  N_z\mathrm{St}$.
Thus, the condition $\tau_{z\mathrm{d}}\lesssim 1$ requires $\mathrm{St}\lesssim N_z^{-1}$.

\subsection{Gas-Dominated Regime}

For the gas-dominated regime ($f_\mathrm{dg}<1$),
there are two important growing modes of SI \citep{YoudinGoodman2005ApJ...620..459Y}.
One is the long-wavelength ridge. 
In the strong-coupling limit ($\mathrm{St}\ll 1$), the most unstable mode exists in 
\begin{equation}
 K_x^2 \sim \frac{(1+f_\mathrm{dg})^3}{\mathrm{St}} K_z.
\end{equation}
The nondimensional stopping time in the $x$-direction is 
\begin{equation}
    \tau_{x\mathrm{d}}  
    = \frac{\mathrm{St}N_x}{2\pi \Pi}\left(\frac{(1+f_\mathrm{dg})^3}{\mathrm{St}} K_z\right)^{1/2}
    \gtrsim 
    \frac{N_x (1+f_\mathrm{dg})^{3/2}}{\sqrt{2\pi\Pi}} \sqrt{\mathrm{St}},
\end{equation}
where 
the $x$ wavelength is resolved by  $N_x$ cells
and $K_z\gtrsim 2\pi \Pi$ is used. 
The condition $\tau_{x\mathrm{d}} \lesssim 1$ can be written as 
\begin{equation}
    \mathrm{St}\lesssim \frac{2\pi \Pi}{N_x^2 (1+f_\mathrm{dg})^3}.
\end{equation}
To satisfy both constraints $\tau_{x\mathrm{d}}\lesssim 1$ and $\tau_{z\mathrm{d}}\lesssim 1$, 
the Stokes parameter should satisfy 
\begin{equation}
    \mathrm{St}\lesssim \min \left( \frac{1}{N_z}, \frac{2\pi \Pi}{N_x^2(1+f_\mathrm{dg})^3}\right).
\end{equation}
For $\Pi \sim 0.05$ and $f_\mathrm{dg}=0.2$, 
$\mathrm{St}$ should be less than 
$3\times 10^{-3}(N_x/8)^{-2}$.
The corresponding growth rate is extremely small since 
the growth rate is proportional to $\Omega\mathrm{St}$.

The other growing mode is the short-wavelength branch where 
$K_x \sim \mathrm{St}^{-1}$ and $K_z > K_x$.
The nondimensional stopping time in the $x$-direction becomes
\begin{equation}
    \tau_{x\mathrm{d}} \sim \frac{N_x}{2\pi \Pi}.
\end{equation}
For a fiducial value of $\Pi\sim 0.05$, $\tau_{x\mathrm{d}} \sim 3N_x$ is always larger than unity.
This means that the short-wavelength branch cannot satisfy $\tau_\mathrm{d} \lesssim 1$.

\subsection{Dust-Dominated Regime}

For the dust-dominated regime ($f_\mathrm{dg}>1$), \citet{YoudinGoodman2005ApJ...620..459Y}
found a rapidly growing branch that 
behaves differently from the unstable modes for $f_\mathrm{dg}<1$.

For $K_x\lesssim K_z$, 
this mode emerges only when 
$K_x\mathrm{St}\gtrsim {\cal F}(f_\mathrm{dg})$
\citep{SquireHopkins2018MNRAS.477.5011S},
where ${\cal F}(f_\mathrm{dg}) > 1$.
We obtain 
\begin{equation}
    \tau_{x\mathrm{d}} >
    25 \left(\frac{N_x}{8}\right)
    \left(\frac{\Pi}{0.05}\right)^{-1},
\end{equation}
where $K_x\mathrm{St}> 1$ is used.
This indicates that the high-$f_\mathrm{dg}$ mode for $K_x\lesssim K_z$
does not grow when $\tau_{x\mathrm{d}}\lesssim 1$.

In the opposite limit $K_x \gg K_z$, 
the maximum growth rate is $(K_z/K_x)\Omega \sqrt{f_\mathrm{dg}-1}$
and the instability occurs only when $K_x\mathrm{St} \gtrsim (K_z/K_x){\cal C}(f_\mathrm{dg})$,
where $\mathcal{C}(f_\mathrm{dg})$ can be approximated as $\sqrt{(f_\mathrm{dg}-1)^{-1} + (0.2f_\mathrm{dg}+0.5)^2}$ 
\citep{SquireHopkins2018MNRAS.477.5011S}.
Existence of the unstable modes satisfying $\tau_{x\mathrm{d}}\lesssim 1$ 
requires 
\begin{equation}
\frac{K_x}{K_z} \gtrsim \frac{N_x \mathcal{C}(f_\mathrm{dg})}{2\pi \Pi}.
\end{equation}
For $\Pi=0.05$, $K_x/K_z \gtrsim 34(N_x/8)$ for $f_\mathrm{dg}= 2$ 
and $K_x/K_z \gtrsim 1.7\times 10^2(N_x/8)$ for $f_\mathrm{dg}=30$.
This condition sets the following upper limit on the maximum growth rate $\sigma$,
\begin{equation}
    \frac{\sigma}{\Omega} \lesssim \frac{2\pi \Pi}{N_x} \frac{\sqrt{f_\mathrm{dg}-1}}{\mathcal{C}(f_\mathrm{dg})}.
\end{equation}
For $\Pi=0.05$, $\sigma/\Omega \lesssim 0.029(N_x/8)^{-1}$ for $f_\mathrm{dg}= 2$ 
and $\sigma/\Omega \lesssim 0.033(N_x/8)^{-1}$ for $f_\mathrm{dg}=30$.
Thus, satisfying $\tau_{x\mathrm{d}}\lesssim 1$ requires a large $K_x/K_z$, leading 
to small maximum growth rate $\propto K_z/K_x$.

Such highly anisotropic modes $(K_x/K_z\gg 1)$ have small growth rates proportional to $K_z/K_x$, and 
their high radial wavenumbers may make them sensitive to physical diffusion and viscosity.

\section{Kelvin-Helmholtz Instability in Multi-fluid Dust}\label{app:kh_multi}

We present a KHI test for a multi-fluid dust-gas system with AMR. 
We implemented the \texttt{Split-Flux} method in the repository provided by \citet{HuangBai2022ApJS..262...11H}.
We use the setup of Section \ref{sec:kh},  except that the magnetic field is removed and 
two dust species, denoted by subscripts 1 and 2, are included.
Following \citet{HuangBai2022ApJS..262...11H}, 
we set $t_\mathrm{stop,d,1} = 10^{-8}$ and $t_\mathrm{stop,d,2}=10^{-2}$.
For species 1, dust--gas drift is unresolved at all AMR levels ($\tau_{\mathrm{d},1}\ll 1$), whereas 
species 2 is marginally coupled with the gas, with $\tau_{\mathrm{d,2}}\gtrsim 1 $ at all AMR levels.
The dust--to--gas mass ratios of species 1 and 2 are unity. 
We conduct the simulations with \texttt{HB22} and \texttt{Split-Flux}.

\begin{figure}[htpb]
    \centering
    \includegraphics[width=1.0\linewidth]{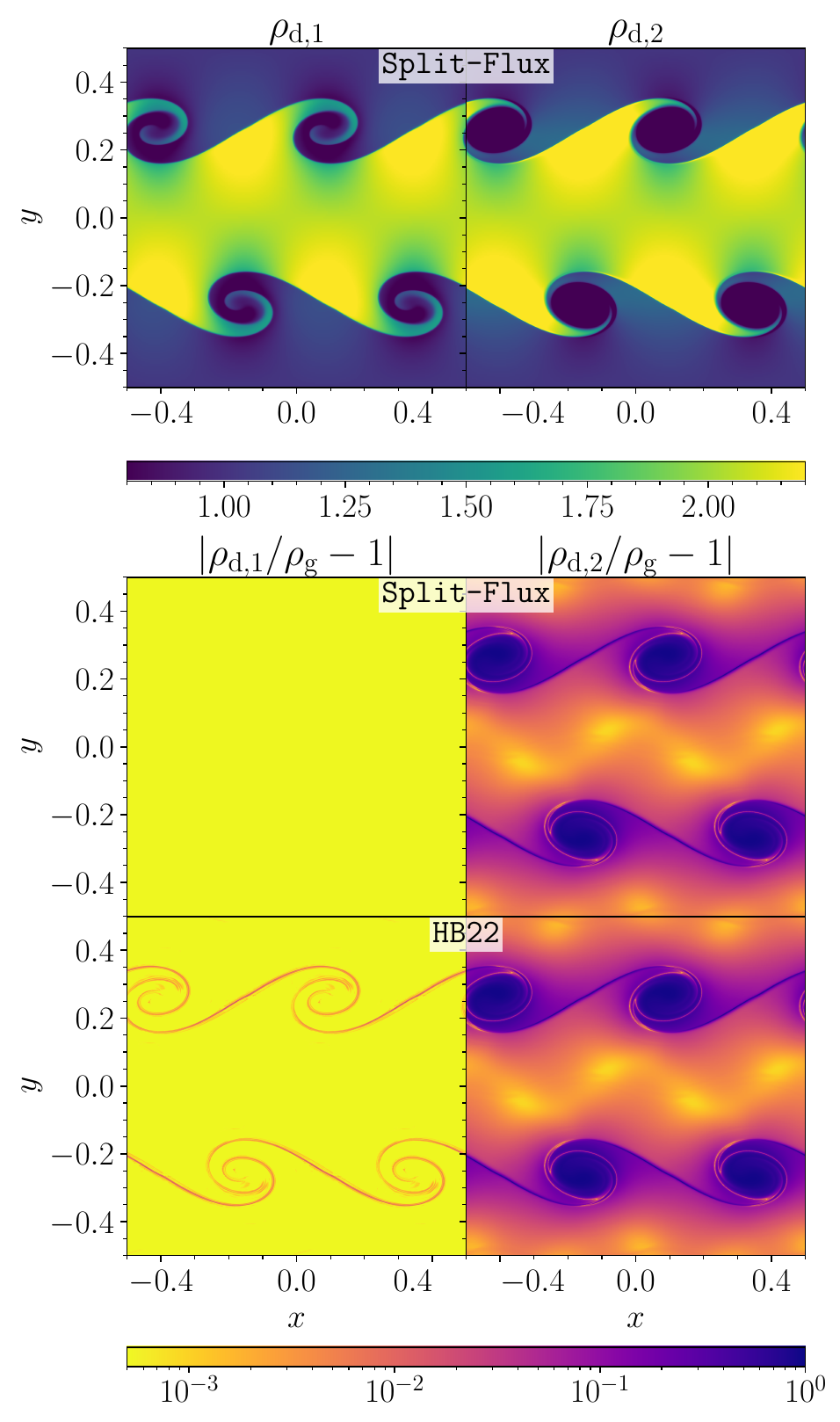}
    \caption{
    Results of the KHI test at $t=1.6$ for \texttt{Split-Flux} and \texttt{HB22} with AMR.
    The first row shows the dust densities of species 1 and 2 obtained with \texttt{Split-Flux}.
    The second and third rows compare, for species 1 (left) and 2 (right), the deviations of the dust--to--gas mass ratios from their initial 
    values in the \texttt{Split-Flux} and \texttt{HB22} runs.
    }
    \label{fig:kh_multi}
\end{figure}

The top row of Figure \ref{fig:kh_multi} compares $\rho_{\mathrm{d,1}}$ and $\rho_\mathrm{d,2}$ 
for \texttt{Split-Flux}.
Dust species 2 is depleted around the vortex centers, consistent with \citet{HuangBai2022ApJS..262...11H}.

The middle and bottom rows of Figure \ref{fig:kh_multi} demonstrate the behavior of the 
\texttt{Split-Flux} method in a multi-fluid calculation.
For species 1, whose dust--gas drift is unresolved, 
the \texttt{Split-Flux} method can keep the dust--to--gas mass ratio constant, whereas 
deviations are seen along the shear layers for \texttt{HB22}.
These two methods yield consistent results for dust species 2 as expected.

\bibliography{ms}{}
\bibliographystyle{aasjournalv7}



\end{document}